\documentclass[aip,amsmath,amssymb,
reprint,
floatfix
]{revtex4-1}
\usepackage{amsfonts,amsmath,bm,multirow}
\usepackage[OT4]{fontenc}

\usepackage{etoolbox}

\usepackage{graphicx}
\usepackage{physics}
\usepackage{hyperref}
\usepackage{cleveref}
\usepackage{dcolumn}
\usepackage{subfigure}

\newcommand{\kp}{$\bm{k}\!\vdot\!\bm{p}\,$}

\graphicspath{{wykresy/}}

\makeatletter
\def\@email#1#2{
 \endgroup
 \patchcmd{\titleblock@produce}
  {\frontmatter@RRAPformat}
  {\frontmatter@RRAPformat{\produce@RRAP{*#1\href{mailto:#2}{#2}}}\frontmatter@RRAPformat}
  {}{}
}
\makeatother

\begin{document}

    \title{Electronic and structural properties of group IV materials and their polytypes}

	\author{Jakub Ziembicki}
	\email{jakub.ziembicki@pwr.edu.pl}
	\affiliation{Department of Semiconductor Materials Engineering, Wroc\l aw  University of Science and Technology, Wybrze\.ze Wyspia\'nskiego 27, 50-370 Wroc{\l}aw, Poland}
	\author{Pawe{\l} Scharoch}
	\affiliation{Department of Semiconductor Materials Engineering, Wroc\l aw  University of Science and Technology, Wybrze\.ze Wyspia\'nskiego 27, 50-370 Wroc{\l}aw, Poland}
	\author{Maciej P. Polak}
	\affiliation{Department of Materials Science and Engineering, University of Wisconsin-Madison, 1509 University Ave., Madison, WI 53706, United States}
	\author{Micha{\l} Wi{\'s}niewski}
	\affiliation{Department of Experimental Physics, Wroc\l aw  University of Science and Technology, Wybrze\.ze Wyspia\'nskiego 27, 50-370 Wroc{\l}aw, Poland}
	\author{Robert Kudrawiec}
	\affiliation{Department of Semiconductor Materials Engineering, Wroc\l aw  University of Science and Technology, Wybrze\.ze Wyspia\'nskiego 27, 50-370 Wroc{\l}aw, Poland}

\date{\today}

\begin{abstract}

Nanotechnology's impact on semiconductor industry advancement, particularly through the engineering of nanostructures like nanowires, opens new possibilities for material functionality due to the tunable physical properties of nanostructures compared to bulk materials. This paper presents a comprehensive study on group IV semiconductors and their binaries across four polytypes: 2H, 3C, 4H, and 6H, focusing on their optoelectronic application potential. Deep understanding of these polytypes is particularly relevant for nanowire-based technologies. Through first principles modeling, we examine the structural and electronic properties of these materials, emphasizing their band structure, stability, and the feasibility for light-emitting applications. We use a generalized Ising model to discuss materials stability and tendency for polytypism. We also determine relative band edge positions and employ a six \kp model for a detailed understanding of the materials' electronic properties. Due to the comprehensive nature of this study, we provide insight on the chemical trends present in all of the studied properties. Our theoretical predictions align well with existing experimental data, suggesting new avenues for nanostructure-based device development. The discussion extends to the implications of these findings for the fabrication of optoelectronic devices with the studied IV-IV materials, highlighting the challenges and opportunities for future research in nanowire synthesis and their application.

\end{abstract}
	
\maketitle

\section{Introduction}

Nanotechnology has become a crucial part of development in the semiconductor industry. It offers new possibilities in engineering of materials functionality, as physical properties of nanostructures can be tuned more freely than those of bulk materials \cite{III-V_NW_review,NW_review,NW_review2,NW_review3,SiGe_NW}. It is, therefore, no surprise that much effort has been put into improving the ability of growing and synthesis of such structures \cite{NanoGrowth,NanoGrowth2}. Among them, nanowires are especially interesting, as they give new possibilities in mixing different materials. They exhibit better tolerance to lattice mismatch, allowing broader range of heterostructures and alloys to be made \cite{Crit_dimension}. Heterostructures within such systems often posses abrupt interfaces \cite{SiGe_NW2,SiGe_NW3} which enhance controllability of its properties. Additionally this growth mode supports formation of hexagonal polytypes \cite{polytypes,polytypes2}, which for most cubic semiconductors of group IV and III-V are unstable in bulk form. Properties of different polytypes of a given material may differ but at the same time they stay lattice matched. This presents new opportunities in tailoring of the properties of these materials for use in novel semiconductor devices.

Among group IV semiconductors the most prominent example of polytypism is present in silicon carbide (SiC), with hundreds of different polytypes reported \cite{SiC_polytypes}. The origin of such broad spectrum of different structures lies in the stacking sequence, where there are unlimited numbers of permutations of possible relative bilayer positions along the [0001] crystallographic direction. Most common forms of SiC are the 2H, 3C, 4H and 6H structures, where the number denotes the amount of bilayers in the primitive cell and the letter stands for cubic (C) or hexagonal (H) crystal symmetry. As SiC is of great importance for the industry due to its mechanical, thermal and electronic properties, its polytypism was a subject of numerous studies \cite{SiC_polytypes,SiC_book,SiC_book2}. SiC nanowires also attracted much attention due to a perspective of combining properties of SiC with advantages of nanostructures \cite{SiC_nanostructures}.

Although in their bulk form, most of group IV and III-V semiconductor do not exhibit polytypism, the progress in techniques of synthesis allows for formation of nanowires composed of different polytypes \cite{polytypes,polytypes2}. Many of such structures offer new opportunities for applications, particularly in optoelectronics. An example may be 2H germanium, which, due to its direct band gap, may be used for light emission, unlike its typical cubic (3C) counterpart \cite{Ge_PL,SiGe_NW}. Although lowest optical transition in pure 2H germanium has small oscillator strength \cite{Ge_OscStr}, it has been shown that by applying strain \cite{Ge_strain,Ge_perturbations} or alloying \cite{Ge_PL,Ge_alloy,Ge_perturbations} it can exhibit strong photoluminescence in near infrared region, at the same time being compatible with silicon technology. Other promising materials for optoelectronics with possible integration with silicon are germanium-tin alloys. It is predicted that germanium alloyed with over 10\% of tin is a direct gap semiconductor with mid infrared emission \cite{GeSn_directGap,GeSn_DirectGap2,GeSn_DirectGap3}, but synthesis of such alloy in bulk turns out to be challenging. On the other hand, nanowires based on germanium core with as much as 10\% of tin have been successfully grown and characterised \cite{GeSn,GeSn2}.

Taking into account the new opportunities opened up by nanowires of group IV semiconductors, broad theoretical studies of their hexagonal polytypes are highly desirable. Predicting their electronic properties from first principles may enhance progress in development of nanostructure-based devices. In this work we address this problem by modeling all group IV elements (excluding lead) and binaries in four most relevant polytypes: 2H, 3C, 4H and 6H. Our studies are mainly focused on optoelectronic applications, therefore we place emphasis on the materials' structure and electronic properties. For better understanding the latter, we derive relative band edge positions and parameterize our band structure results with six \kp model, which is broadly used in modeling III-V nitrides in 2H phase \cite{kp1,kp2,nextnano}. Our studies also include stability considerations to suggest which materials are technologically attainable and which polytype is energetically preferred. We discuss possible candidates for light emitting applications, taking into account their band gaps and selection rules. Our predictions are in good agreement with existing theoretical and experimental studies.

This paper is arranged as follows: in Sec. \ref{Ch:results}, the results of our calculations are described and are divided into structural properties, stability considerations, electronic structure properties, and relative band edge positions. In Sec. \ref{Ch:discussion} we discuss materials which are promising for optoelectronic applications. In Sec. \ref{Ch:conclusion} we summarize our studies, and, finally, we describe the used methods in Sec. \ref{Ch:method}.
 
\section{Results}\label{Ch:results}

\subsection{Structural properties}\label{Ch:structural_prop}

\begin{figure*}
    \centering
    \includegraphics[width=1\textwidth]{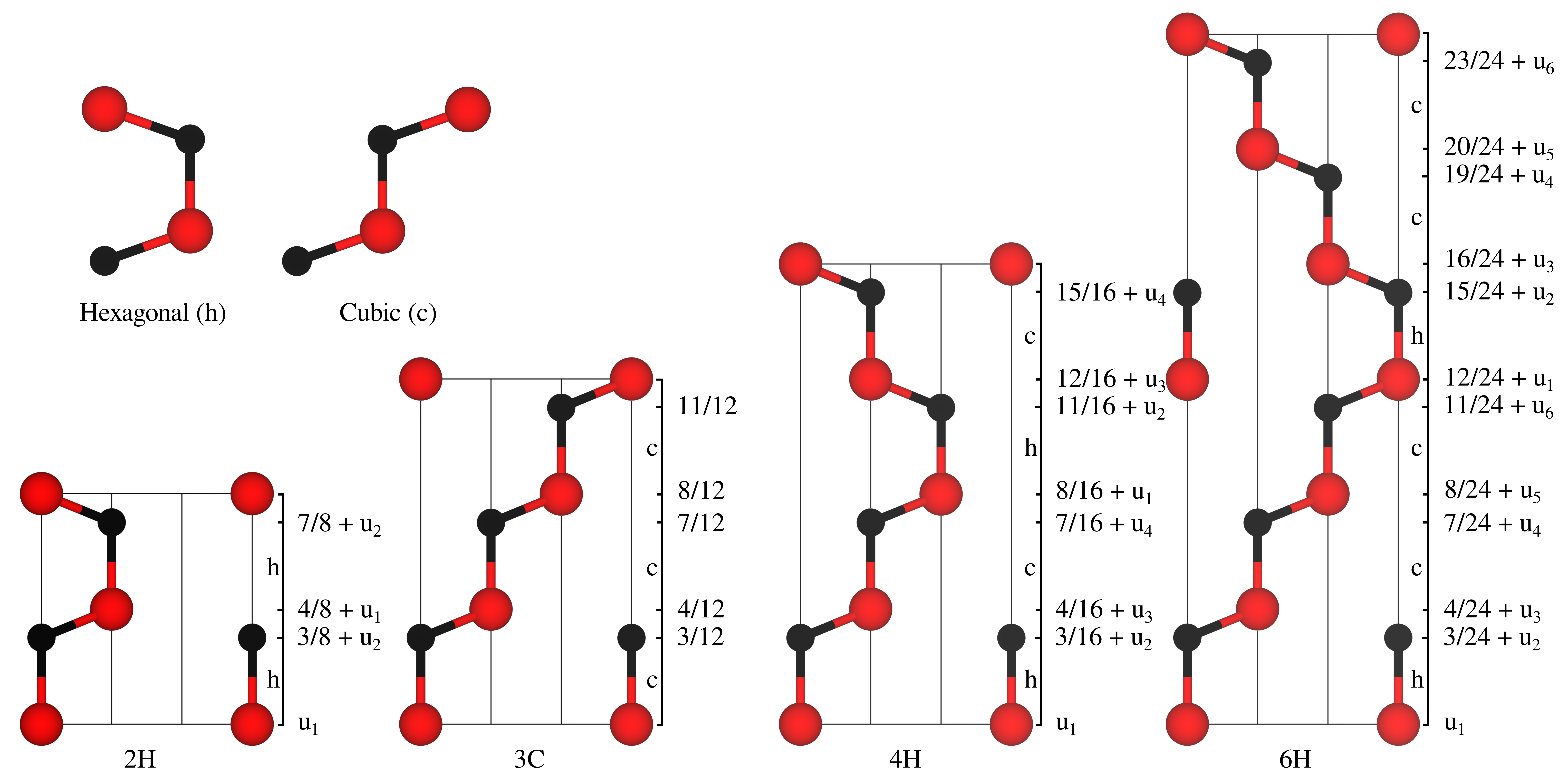}
\caption{Hexagonal unit cell (supercell in the  3C case) of studied polytypes. The (11\={2}0) plane is presented and bonds in this plane are marked. Different stacking of hexagonal and cubic bilayers differentiate polytypes. Whether a bilayer is hexagonal (h) or cubic (c) is determined by the limiting bonds being either parallel (c) or not parallel (h).  Internal cell $u_i$ parameters are also defined.} 
    \label{fig:cell2}
\end{figure*}

\begin{table}
\caption{\label{Tab:lattice} Lattice parameters of studied polytypes.}
\begin{ruledtabular}
\begin{tabular}{ccccc}
     &    & a     & 2c/p  & $B_0$\\ \hline
C    & 3C & 2.499 & 4.081 & 468 \\
     & 6H & 2.494 & 4.100 & 472 \\
     & 4H & 2.491 & 4.110 & 471 \\
     & 2H & 2.485 & 4.139 & 466 \\
Si   & 3C & 3.820 & 6.239 & 97  \\
     & 6H & 3.814 & 6.253 & 97  \\
     & 4H & 3.811 & 6.261 & 97  \\
     & 2H & 3.803 & 6.291 & 97  \\
Ge   & 3C & 3.993 & 6.520 & 72  \\
     & 6H & 3.986 & 6.533 & 72  \\
     & 4H & 3.983 & 6.540 & 72  \\
     & 2H & 3.976 & 6.564 & 73  \\
Sn   & 3C & 4.581 & 7.480 & 45  \\
     & 6H & 4.574 & 7.488 & 46  \\
     & 4H & 4.572 & 7.495 & 46  \\
     & 2H & 4.565 & 7.513 & 46  \\ \hline
CSi  & 3C & 3.062 & 5.000 & 229 \\
     & 6H & 3.060 & 5.005 & 229 \\
     & 4H & 3.060 & 5.008 & 229 \\
     & 2H & 3.057 & 5.018 & 230 \\
CGe  & 3C & 3.223 & 5.262 & 197 \\
     & 6H & 3.219 & 5.270 & 198 \\
     & 4H & 3.217 & 5.274 & 198 \\
     & 2H & 3.213 & 5.288 & 198 \\
CSn  & 3C & 3.545 & 5.789 & 145 \\
     & 6H & 3.543 & 5.790 & 145 \\
     & 4H & 3.542 & 5.790 & 145 \\
     & 2H & 3.539 & 5.794 & 146 \\
SiGe & 3C & 3.901 & 6.371 & 85  \\
     & 6H & 3.895 & 6.384 & 85  \\
     & 4H & 3.892 & 6.391 & 85  \\
     & 2H & 3.884 & 6.418 & 85  \\
SiSn & 3C & 4.221 & 6.893 & 65  \\
     & 6H & 4.215 & 6.904 & 65  \\
     & 4H & 4.212 & 6.911 & 65  \\
     & 2H & 4.204 & 6.934 & 65  \\
GeSn & 3C & 4.293 & 7.011 & 57  \\
     & 6H & 4.287 & 7.020 & 57  \\
     & 4H & 4.284 & 7.026 & 57  \\
     & 2H & 4.277 & 7.047 & 57 
\end{tabular}
\end{ruledtabular}
\end{table}

\begin{table*}
\caption{\label{Tab:ui}  Internal cell parameters  $u_i$  (see Fig. \ref{fig:cell2}) ($\times 10^{-4}$).}
\begin{ruledtabular}
\begin{tabular}{cccccccccc}
     & $u_2^{2H}$      & $u_2^{4H}$     & $u_3^{4H}$     & $u_4^{4H}$     & $u_2^{6H}$     & $u_3^{6H}$     & $u_4^{6H}$    & $u_5^{6H}$     & $u_6^{6H}$     \\ \hline
C    & -6.757  & 13.975 & 15.087 & -1.115 & 12.674 & 14.004 & 6.975 & 5.699  & -1.319 \\
Si   & -10.916 & 6.512  & 8.462  & -1.942 & 6.363  & 7.711  & 3.445 & 2.917  & -1.349 \\
Ge   & -8.150  & 7.638  & 9.556  & -1.913 & 6.950  & 8.523  & 3.763 & 3.187  & -1.563 \\
Sn   & -4.970  & 6.671  & 8.043  & -1.382 & 6.020  & 6.993  & 3.213 & 2.797  & -0.983 \\ \hline
CSi  & 8.010   & 5.362  & 6.412  & 5.881  & 3.565  & 0.722  & 0.592 & -1.078 & -1.202 \\
CGe  & -3.918  & 4.091  & 6.980  & 1.421  & 3.641  & 3.093  & 0.988 & 0.386  & -2.060 \\
CSn  & 5.362   & 3.480  & 4.320  & 3.767  & 2.188  & 0.062  & 0.114 & -0.518 & -0.798 \\
SiGe & -10.050 & 6.604  & 8.066  & -2.604 & 6.305  & 8.207  & 3.673 & 3.053  & -1.154 \\
SiSn & -9.188  & 5.353  & 8.127  & -1.223 & 5.160  & 6.022  & 2.242 & 1.945  & -1.961 \\
GeSn & -6.630  & 6.719  & 9.453  & -0.444 & 6.128  & 6.452  & 2.665 & 2.245  & -1.924
\end{tabular}
\end{ruledtabular}
\end{table*}

Structures of studied polytypes are presented in Fig. \ref{fig:cell2} in hexagonal unit cells, where the bonds lying in the crystallographic (11\={2}0) plane are shown. For easier comparison of different phases the cubic structure is represented in the  3C  hexagonal supercell. One can note that the difference between phases arise from stacking sequences of atomic bilayers along the [0001] crystallographic direction, which are ABA, ABCA, ABCBA and ABCACBA for 2H, 3C 4H and 6H respectively. Apart from that, the bilayers are classified into two classes: hexagonal (h), when the limiting bonds are not parallel, and cubic (c), when the limiting bonds are parallel. Equivalently, for hexagonal bilayers one of the limiting bonds is rotated by $180^{\circ}$ with respect to the other around chemical bond along [0001], and for cubic bilayers it is not rotated. One can note in Fig. \ref{fig:cell2} that the phases differ in the number of $h$ and $c$ bilayers. Based on this observation a parameter called the level of hexagonality as the ratio of hexagonal layers to the total number of layers in unit cell can be defined. The levels of hexagonality for 2H, 3C, 4H and 6H are respectively: $1$, $0$, $\frac{1}{2}$ and $\frac{1}{3}$.

A formal description of the geometry is given by two lattice parameters $a$, $c$ and $(p-1)$ internal cell parameters $u_i$, which establish degrees of freedom in atomic positions. Our choice of unit cells are the following basis vectors: 
\begin{equation*}
    a_1=a(1,0,0),\quad a_2=a(-\frac{1}{2},\frac{\sqrt{3}}{2},0),\quad a_3=c(0,0,1)
\end{equation*}
and atomic positions in reduced coordinates: 
\begin{equation*}
    (0,0,z),\quad  (\frac{1}{3},\frac{2}{3},z),\quad  (\frac{2}{3},\frac{1}{3},z), 
\end{equation*}
where $z$ for each atom is defined in Fig. \ref{fig:cell2} by means of $u_i$ parameters, whose values are given in Tab. \ref{Tab:ui}. 

The nearest neighbors configuration in each structure is a tetrahedron, which is slightly distorted for hexagonal phases (so that the bond lengths and angles are not all the same). For an ideal structure $\frac{2c}{pa}=\sqrt{\frac{8}{3}}$ and $u_i=0$, which is the case for 3C, whereas for hexagonal crystals these equations in general do not hold. The numerical results of geometrical parameters from our calculations are presented in Tab. \ref{Tab:lattice} and Tab. \ref{Tab:ui}. As it can be seen in Fig. \ref{fig:EneCratioJ} in all investigated materials the $\frac{2c}{pa}$ ratios are greater than the ideal value and increase linearly with hexagonality, which is in agreement with previous studies \cite{SiGe}. It is however in contrast with earlier empirical rule that when material is stable in the 2H phase then this ratio is smaller than the ideal one \cite{stability,stability2}, which is not the case for CSn. Separately, the lattice constant $a$ always decreases with hexagonality and the ratio $\frac{2c}{p}$ increases, although in general different polytypes are rather lattice matched, with differences in the $a$ parameter not exceeding 0.056\%. On the other hand, bulk modulus which quantify the material's elasticity are almost constant for all values of hexagonality. Small differences in bulk modulus and in plane lattice constant indicate that polytypes can freely coexist in both strained and unstrained heterostructures.

\subsection{Stability of polytypes}\label{Ch:stability}

\begin{table}
\caption{\label{Tab:Ji}  Parameters $J_i$ of ANNNI model, in meV per bilayer pair.}
\begin{ruledtabular}
\begin{tabular}{cccc}
     & $J_1$     & $J_2$     & $J_3$\\ \hline
C    & 25.448 & -3.435 & -0.337 \\
Si   & 10.587 & -2.593 & -0.766 \\
Ge   & 16.816 & -1.054 & -0.344 \\
Sn   & 15.178 & -0.222 & -0.045 \\ \hline
CSi& 2.735  & -2.424 & -0.396 \\
CGe& 3.032  & -1.752 & -0.335 \\
CSn& -4.002 & -1.120 & -0.233 \\
SiGe & 12.586 & -1.901 & -0.583 \\
SiSn & 11.125 & -1.174 & -0.353 \\
GeSn   & 15.336 & -0.528 & -0.180
\end{tabular}
\end{ruledtabular}
\end{table}

\begin{figure*}
    \centering
    \includegraphics[width=1\textwidth]{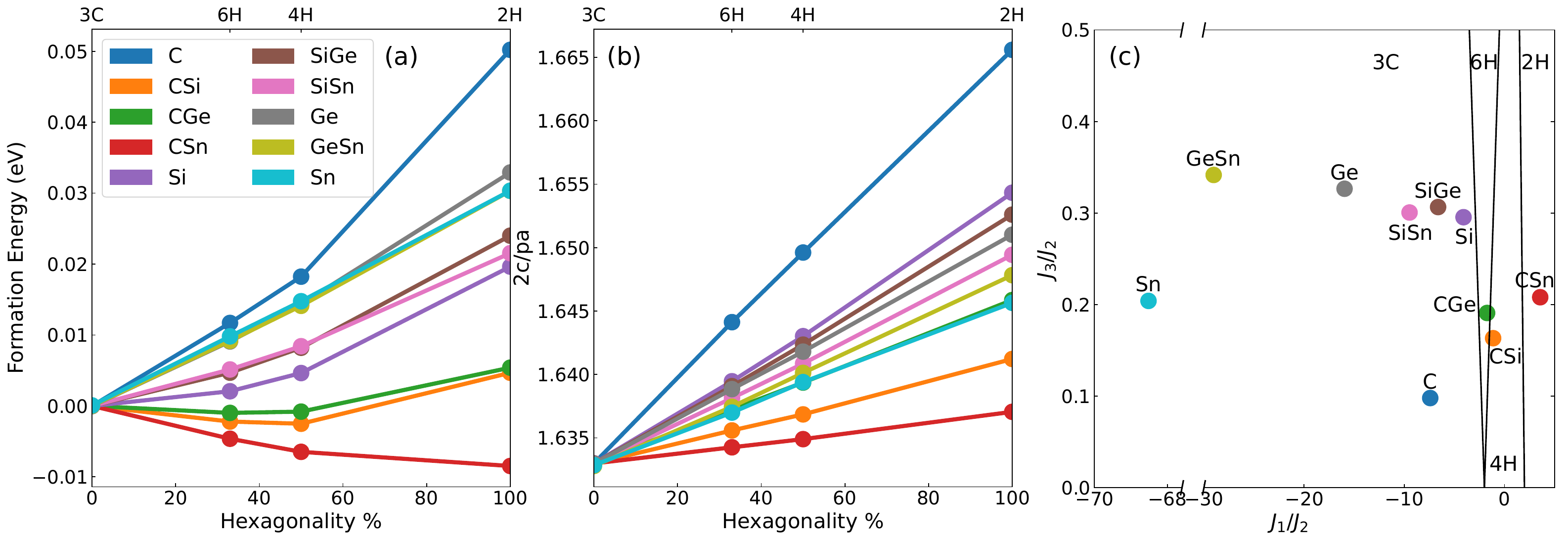}
    \caption{Energetic and geometrical trends in function of hexagonality: formation energy per atom pair (a) and $\frac{2c}{pa}$ ratio (b). Panel (c) represent phase diagram of ANNNI model with phase boundaries marked by black solid lines.} 
    \label{fig:EneCratioJ}
\end{figure*}

The formation energy of the crystal relative to 3C phase is presented in Fig. \ref{fig:EneCratioJ} (numerical values are presented in Tab. 1 in supplementary materials). It is clear that for all materials except binaries with carbon 3C phase is energetically favored. For CSi and CGe 4H and 6H phases have lowest energies, relative close to each other, whereas CSn has minimum at 2H phase. For the rest of materials a trend of increase in energy with hexagonality is observed. Similar trends for Si, Ge and CSi were predicted by previous theoretical studies \cite{SiC_energy,SiGe}. It seems that the presence of carbon in group IV binaries enhance stability of hexagonal polytypes. It should be stressed out, that CGe, CSn, SiGe, SiSn and GeSn binaries are predicted to be unstable at zero temperature and zero pressure, with the positive formation energy of 0.408, 0.784, 0.022, 0.155, 0.029 eV/atom, respectively. Also it is known that the energetically stable phase of carbon is the graphite phase, although the 3C phase is also effectively stable due to long phase transition time.

To gain better insight into the stability of polytypes, we adopted a previously proposed axial nearest neighbor Ising (ANNNI) model \cite{ANNNI_SiC1,ANNNI_SiC2,III-VRev} to study polytypism based on total energy of the system from DFT calculations. It is an Ising model where spin up $\sigma\uparrow$ and spin down $\sigma\downarrow$ are replaced by bonds direction of adjacent bilayers, (bonds on Fig. \ref{fig:cell2} not parallel to [0001] direction). The hamiltonian describing the energy part of the system originating from different stacking takes the form:
\begin{equation*}
   E=-\frac{1}{p}\sum^p_{i=1}\sum^\infty_{j=1}J_j\sigma_i\sigma_{i+j}
\end{equation*}
where the summation over i includes all bilayers in unit cell of a $p$H polytype. Summation over $j$ runs over all bilayers in crystal and we truncate it to three nearest bilayers. The  model contains three parameters, $J_1$, $J_2$ and $J_3$, which can be calculated from formation energy difference of different polytypes (formulas can be found elsewhere \cite{III-VRev}). Results of our calculations are given in Tab. \ref{Tab:Ji}. It can be observed that in terms of absolute values for most materials $J_1$ is an order of magnitude bigger than $J_2$ and similarly $J_2$ is bigger than $J_3$. Another difference is in that $J_1$ has positive sign whereas $J_2$ and $J_3$ are negative. Large positive values of $J_1$ for most materials are the reason for 3C phase being most stable. The exceptions are CSi, CGe and CSn, where $J_1$ and $J_2$ have similar absolute values. CSn is the only material with a negative $J_1$, which is why the 2H phase is energetically favored. Similar values of $J_1$ and $J_2$ parameters indicate that material has stable hexagonal polytypes and the sign of $\frac{J_1}{J_2}$ determines whether it is a 6H/4H (minus) or 4H/2H (plus) phase. 

The above discussion can be illustrated by the phase diagram in Fig. \ref{fig:EneCratioJ}. It can be seen that materials that are close to the phase diagram boundaries are these with similar values of $J_1$ and $J_2$. Among all materials, only CSi and CGe are in the region which suggests the strongest polytypism (which for CSi is experimentally known) and CSn is the only material which prefers the 2H phase. Apart from the materials mentioned above, two materials are relatively close to 3C/6H boundary, Si and SiGe, which indicates that these materials may prefer hexagonal phase in nanostructure growth.

\subsection{Electronic band structure}\label{Ch:bs}

\begin{table}
\caption{\label{Tab:EgDeltas} Fundamental band gaps and $\Delta_{cr}$, $\Delta_{so}$ splittings. Positions of conduction band minima in the Brillouin zone (points of high symmetry) are given next to band gap values. Asterisks denote that minimum is not precisely in a high symmetry point but close to it. Branch point energy is also given with respect to valence band maximum.}
\begin{ruledtabular}
\begin{tabular}{cccccc}
     &    & $Eg$ (eV)       & $\Delta_{cr}$ (eV)  & $\Delta_{so}$ (eV) & $E_{bp}$ (eV)  \\ \hline
C    & 3C & 4.987 X*  & 0.000 & 0.012 & 1.532 \\
     & 6H & 5.264 M*  & 0.157 & 0.012 & 1.484 \\
     & 4H & 5.381 M*  & 0.234 & 0.012 & 1.369 \\
     & 2H & 4.400 K    & 0.512 & 0.013 & 1.076 \\
Si   & 3C & 1.219 X*  & 0.000  & 0.047 & 0.274 \\
     & 6H & 1.218 M*  & 0.109 & 0.048 & 0.217 \\
     & 4H & 1.201 M*  & 0.164 & 0.049 & 0.146 \\
     & 2H & 1.058 M  & 0.361 & 0.050 & -0.062  \\
Ge   & 3C & 0.767 L  & 0.000  & 0.277 & -0.114 \\
     & 6H & 0.641 $\Gamma$  & 0.096 & 0.278 & -0.165 \\
     & 4H & 0.589 $\Gamma$  & 0.143 & 0.277 & -0.226 \\
     & 2H & 0.408 $\Gamma$  & 0.286 & 0.286 & -0.385 \\
Sn   & 3C & -0.434 $\Gamma$ & -      & - & -      \\
     & 6H & -0.257 $\Gamma$ & -      & - & -      \\
     & 4H & -0.253 $\Gamma$ & -      & - & -      \\
     & 2H & -0.190 $\Gamma$ & -      & - & -      \\ \hline
CSi  & 3C & 2.330 X   & 0.000  & 0.014 & 1.521 \\
     & 6H & 2.958 M  & 0.036 & 0.014 & 1.540 \\
     & 4H & 3.179 M  & 0.053 & 0.014 & 1.471 \\
     & 2H & 3.298 K  & 0.119 & 0.014 & 1.433 \\
CGe  & 3C & 2.426 X  & 0.000 & 0.039 & 1.169 \\
     & 6H & 2.890 L*   & 0.058 & 0.035 & 1.173 \\
     & 4H & 2.974 M*  & 0.084 & 0.035 & 1.109 \\
     & 2H & 3.251 M  & 0.171 & 0.033 & 1.008 \\
CSn  & 3C & 1.477 $\Gamma$  & 0.000 & 0.061 & 0.796 \\
     & 6H & 1.543 $\Gamma$  & 0.024 & 0.063 & 0.838 \\
     & 4H & 1.582 $\Gamma$  & 0.033 & 0.067 & 0.805 \\
     & 2H & 1.690 $\Gamma$   & 0.096 & 0.042 & 0.772 \\
SiGe & 3C & 1.120 X*   & 0.000 & 0.166 & 0.103 \\
     & 6H & 1.081 M*  & 0.104 & 0.166 & 0.051 \\
     & 4H & 1.092 M*  & 0.160  & 0.160 & -0.017  \\
     & 2H & 0.997 M  & 0.314 & 0.182 & -0.203 \\
SiSn & 3C & 1.012 L  & 0.000 & 0.300 & -0.008   \\
     & 6H & 0.889 $\Gamma$  & 0.078 & 0.302 & -0.043 \\
     & 4H & 0.842 $\Gamma$  & 0.117 & 0.302 & -0.096 \\
     & 2H & 0.678 $\Gamma$  & 0.238 & 0.306 & -0.234 \\
GeSn & 3C & -0.028 $\Gamma$ & -      & - & -     \\
     & 6H & -0.030 $\Gamma$ & -      & - & -      \\
     & 4H & -0.030 $\Gamma$ & -      & - & -      \\
     & 2H & -0.055 $\Gamma$ & -      & - & -     
\end{tabular}
\end{ruledtabular}
\end{table}

\begin{figure*}
    \centering
    \includegraphics[width=0.91\textwidth]{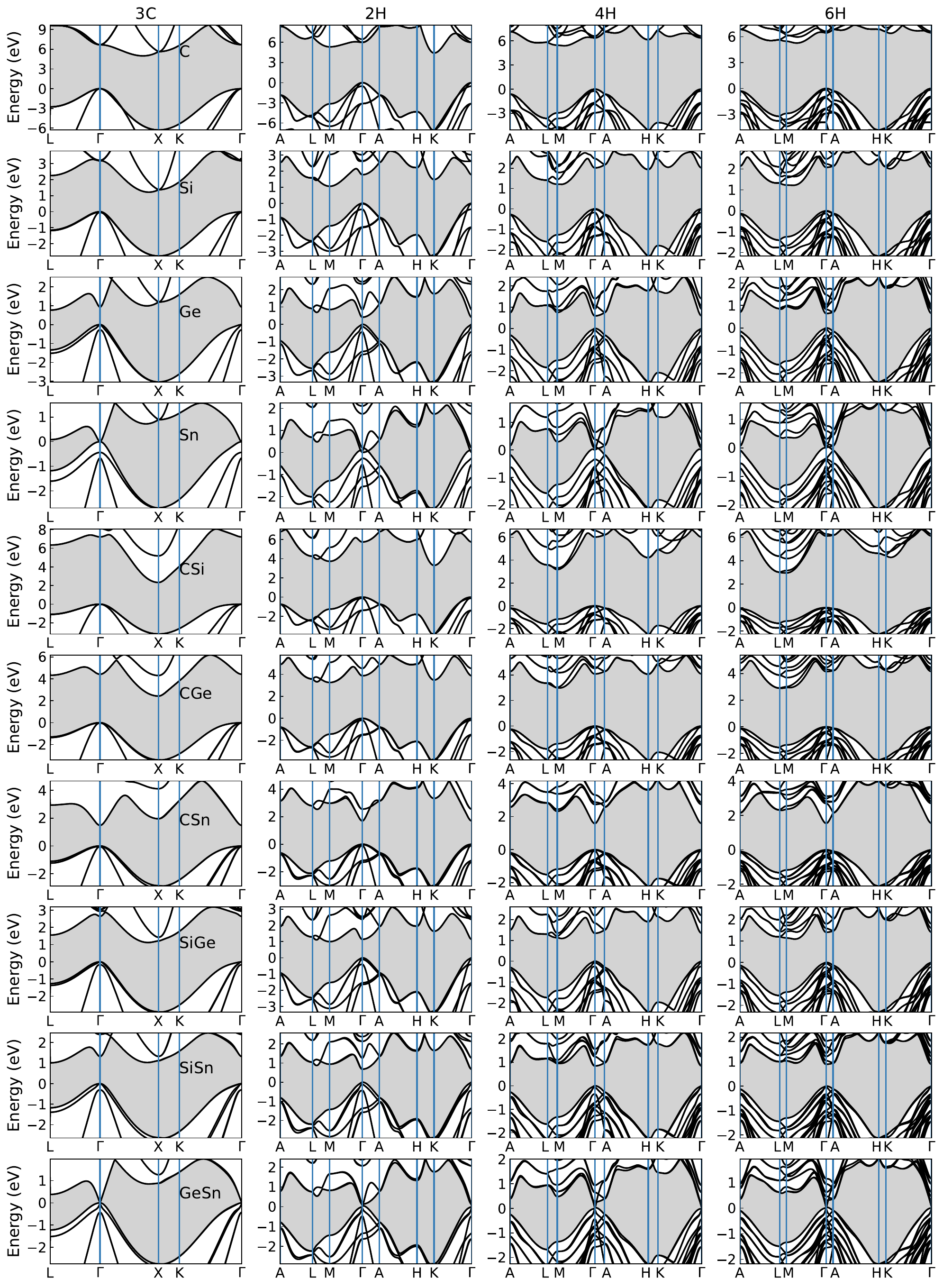}
    \caption{Band structures of the studied IV-IV materials. Different polytypes are in different columns, different chemical compositions are in different rows.} 
    \label{fig:BS}
\end{figure*}

Electronic band structures of studied materials are presented in Fig. \ref{fig:BS} and their most important parameters in Tab. \ref{Tab:EgDeltas}. Due to similar chemical environment of atoms, structures are qualitatively similar. Many differences can be explained on the grounds of bands folding which appears when the size of elementary cell and the number of atoms increase, leaving nearest neighbors configuration of atoms similar. In such cases, band structures are composed of larger number of bands with increasing degree of folding. It is especially true for $\Gamma-L$ in 3C and $\Gamma-A$ in $p$H, where the bands in the former folds to $p$ bands in the latter. For all studied cases the valence bands maximum is always in the $\Gamma$ point and the three topmost bands are in the same order i.e. $\Gamma_{9v}, \Gamma_{7v+}, \Gamma_{7v}$ (in Koster at el. notation \cite{Koster}). On the other hand, conduction bands minima in different materials are located in different points of Brillouin zone and an ordering of bands also differ.

\begin{figure*}
    \centering
    \includegraphics[width=1\textwidth]{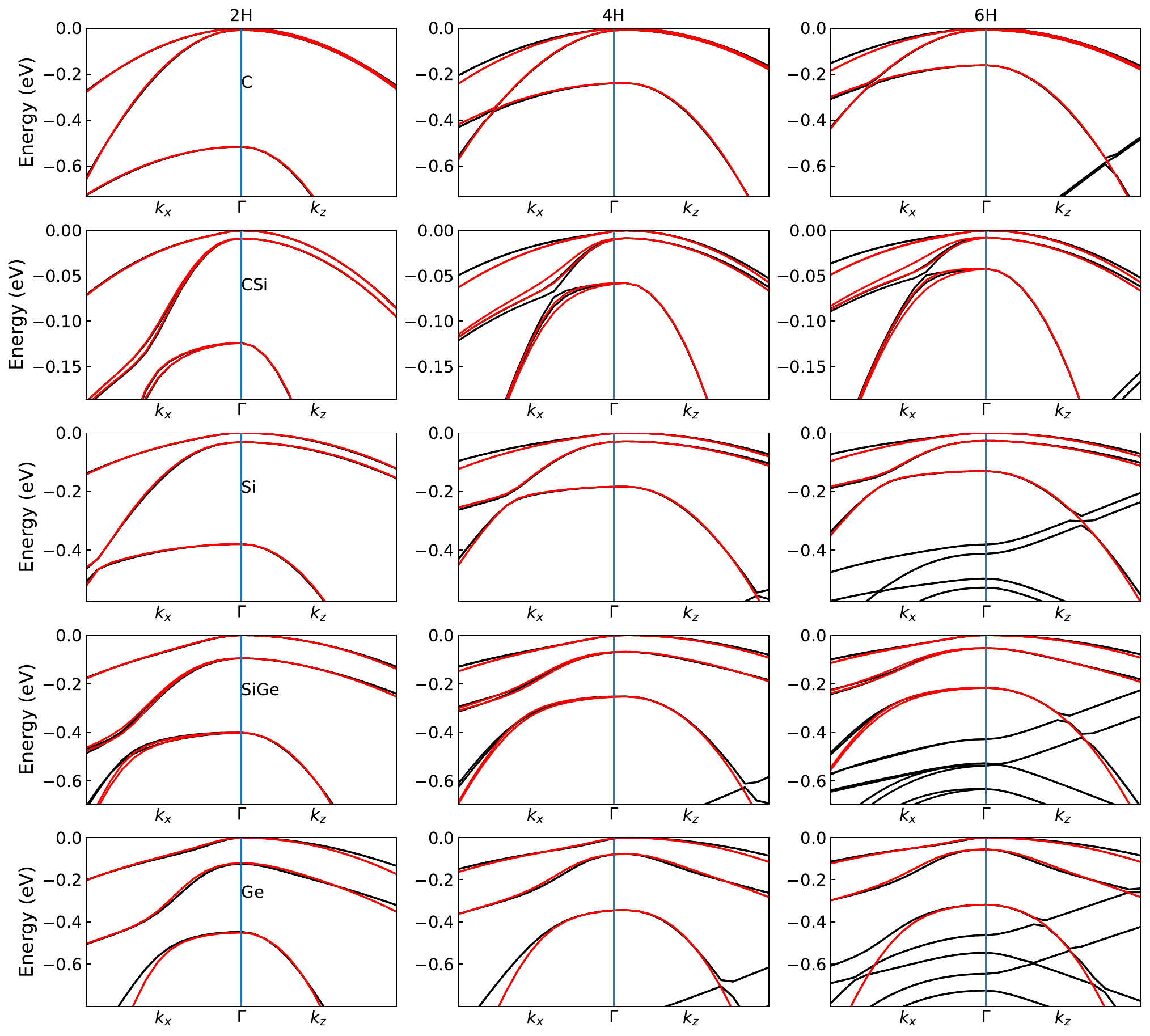}
    \caption{Valence bands of studied materials from DFT (black) and fitted by six \kp model (red). Polytypes change by columns and chemical compositions by rows.}
    \label{fig:kp}
\end{figure*}

\begin{figure*}
    \centering
    \includegraphics[width=1\textwidth]{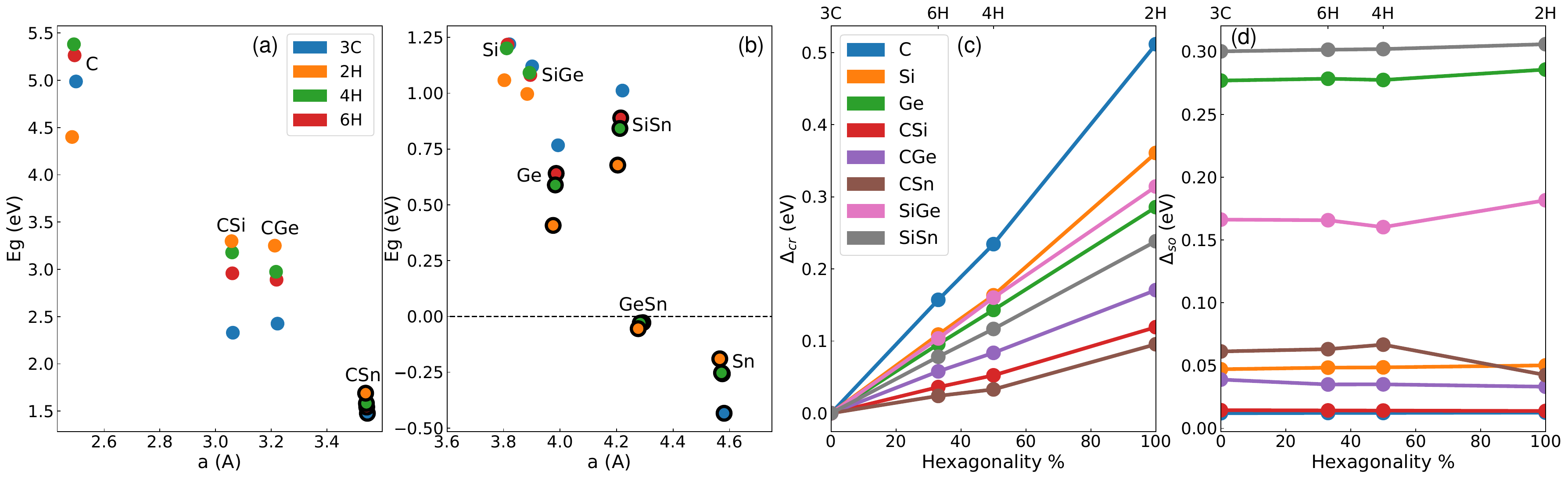}
    \caption{Band gaps and valence bands splittings. Fundamental energy gap vs. in-plane lattice constant is presented on panel (a) for materials with carbon, and on panel (b) for the rest of materials. Materials with direct band gap are marked with a black circle. On panels (c) and (d), splittings $\Delta_{cr}$ and $\Delta_{so}$ are shown as functions of hexagonality.} 
    \label{fig:EgDeltas}
\end{figure*}

There are two visible effects which result from reduction of symmetry. The first is the lifting of degeneracy of light holes and heavy holes at $\Gamma$ point when one change symmetry from cubic to hexagonal. In the hexagonal symmetry, the level of degeneracy is decreased and, as a result, two values of splitting parameters are needed $\Delta_{cr}$ and $\Delta_{so}$ which describe the  energy splitting between bands at the $\Gamma$ point via equation:
\begin{equation*}
    E_{v1}-E_{v2,v3}=\frac{\Delta_{cr}+\Delta_{so}}{2}\mp\frac{\sqrt{\Delta_{cr}^2+\Delta_{so}^2-\frac{8}{3}\Delta_{cr}\Delta_{so}}}{2},
\end{equation*}
where $E_{v1,v2,v3}$ denote subsequent valence band energies. The values of splitting as a function of hexagonality are shown in Fig. \ref{fig:EgDeltas}. It can be seen that values of $\Delta_{so}$ are almost constant among different phases whereas $\Delta_{cr}$ is increasing linearly with hexagonality. One can also observe the well known trend of increasing $\Delta_{so}$ as the atomic mass of elements in materials increase. The other effect related to the change of symmetry can be observed when one changes the material from monoatomic to binary, as these differ by the lack of inversion symmetry in the latter. Materials without inversion symmetry exhibit lifting of spin degeneracy induced by non-zero \textbf{k}-vector in plane perpendicular to [0001] ([111] in 3C) direction because of spin orbit interaction. For the 3C phase, it is a result of the Dresselhaus effect whereas for $p$H phases it is a combination of Dresselhaus and Rashba effects. This effect is most distinct in valence bands of $p$H phases (Fig. \ref{fig:kp}), where, for binaries, lifting of spin degeneracy can be observed along $k_x$ direction.

A closer inspection of valence bands maxima at the $\Gamma$ point reveals that dispersion of bands is very similar across different polytypes, indicating that effective masses do not change significantly. On the other hand, due to complexity of valence bands dispersion near the $\Gamma$ point, they can not be described by simple parabolic model. For this purpose we chose a six \kp model, which was tailored for accurate description of valence band dispersion of III-V nitrides in the 2H phase \cite{kp1,kp2}. Similarities between valence bands at the $\Gamma$ point for different phases suggest that they can be described by a common model with one set of parameters. The six \kp model for the 2H structure contains seven parameters, $A_1$-$A_7$, to describe bands dispersion and two splitting parameters mentioned earlier $\Delta_{so}$, $\Delta_{cr}$. As shown in Fig. \ref{fig:kp}, all hexagonal polytypes are described rather well by the same set of parameters (given in Tab. \ref{Tab:Ai}) with the only difference being in the values of $\Delta_{cr}$ and $\Delta_{so}$. At the same time the parameters from hexagonal phases do not transfer well to the 3C phase, which may be due to a higher symmetry of 3C phase and differences in its geometry, i.e. $u_i$ values and lattice constants ratio. Comparison of DFT and six \kp for all materials and polytypes are presented in Fig. 1 in supplementary data.

As mentioned earlier, conduction bands exhibit larger differences between different materials and phases. Position of the global minimum in Brillouin zone and the character of the band gap (direct or indirect) can change between polytypes. Also the order of bands can change from one material to the other, a good example of which is Ge. Folding of bands cause Ge, which has an indirect gap in 3C with conduction band minimum at L point, to become a direct gap semiconductor in hexagonal phases as the L point folds to $\Gamma$ \cite{Ge_OscStr}. This implies that the lowest conduction band at $\Gamma$ in the 3C phase is the second band in $p$H phases. Using a double group notation the first band has $\Gamma_{8c}$ and the second $\Gamma_{7c}$ symmetry, so the order is reversed in comparison to direct gap 2H semiconductors like GaN. Band ordering is important from the point of view of selection rules. It is known fact that 2H phase of Ge does not exhibit strong photoluminescence, despite being a direct gap semiconductor. The reason is that $\Gamma_{9v}-\Gamma_{8c}$ transition has a small oscillator strength. On the other hand the $\Gamma_{9v}-\Gamma_{7c}$ transition has a large oscillator strength and is optically active, which is the case for nitrides in III-V semiconductors which have $\Gamma_{7c}/\Gamma_{8c}$ bands ordering. The calculated interband momentum matrix elements between first valence band and two lowest conduction bands in 2H are given in Fig. 2 in supplementary data and based on this data one can conclude that from studied materials CSn got $\Gamma_{7c}/\Gamma_{8c}$ ordering whereas Ge, CSi, CGe, SiGe and SiSn got $\Gamma_{8c}/\Gamma_{7c}$ ordering. In the case of C and Si bands ordering is further altered by $\Gamma_{9c}$ states\cite{C_Si_Ge,Si,Si_Ge}.

In Fig. \ref{fig:EgDeltas} the band gap vs. in-plane lattice constant is presented. The almost linear decrease of energy gap with increasing lattice constant can be clearly observed. The values and character of the gap depend on hexagonality. For binaries with carbon and for Sn, the gaps increase with hexagonality, whereas for the rest of materials they decrease. Carbon is an exception, with non-monotonic behaviour. The only materials with direct energy gaps are CSn in all phases, and Ge and SiSn in hexagonal phases. Sn and GeSn are semimetals for all polytypes, with the gap closed at $\Gamma$ point. The rest of the materials have indirect gaps with conduction band minimum either at or next to M, K or L points (X or L for 3C). Because of the previously mentioned folding of the L point in transition from 3C to $p$H, materials with conduction band minima at the L point in 3C (Ge, SiSn) have direct gaps in $p$H.

\begin{table}
\caption{\label{Tab:Ai} $A_i$ parameters of six \kp model (A7 in eV\AA).}
\begin{ruledtabular}
\begin{tabular}{cccccccc}
     & $A_1$      & $A_2$     & $A_3$     & $A_4$     & $A_5$     & $A_6$      & $A_7$      \\ \hline
C    & -5.274  & -1.072 & 3.932  & -1.316 & -0.977 & -0.513  & 0.000  \\
Si   & -8.142  & -1.029 & 6.676  & -2.738 & -2.243 & -2.509  & 0.000  \\
Ge   & -19.529 & -1.390 & 17.253 & -8.391 & -7.897 & -10.732 & 0.000  \\ \hline
CSi  & -4.579  & -0.584 & 3.914  & -1.229 & -1.289 & -1.372  & 0.159  \\
CGe  & -6.005  & -0.793 & 5.208  & -1.703 & -1.725 & -2.015  & 0.148  \\
CSn  & -8.799  & -0.847 & 7.898  & -2.909 & -3.028 & -4.133  & 0.296 \\
SiGe & -11.446 & -1.238 & 9.705  & -4.102 & -3.657 & -4.795  & 0.077  \\
SiSn & -14.589 & -1.526 & 12.578 & -5.236 & -5.041 & -6.804  & 0.154 
\end{tabular}
\end{ruledtabular}
\end{table}

\subsection{Relative band edge positions}\label{Ch:offsets}

\begin{figure*}
    \centering
    \includegraphics[width=1\textwidth]{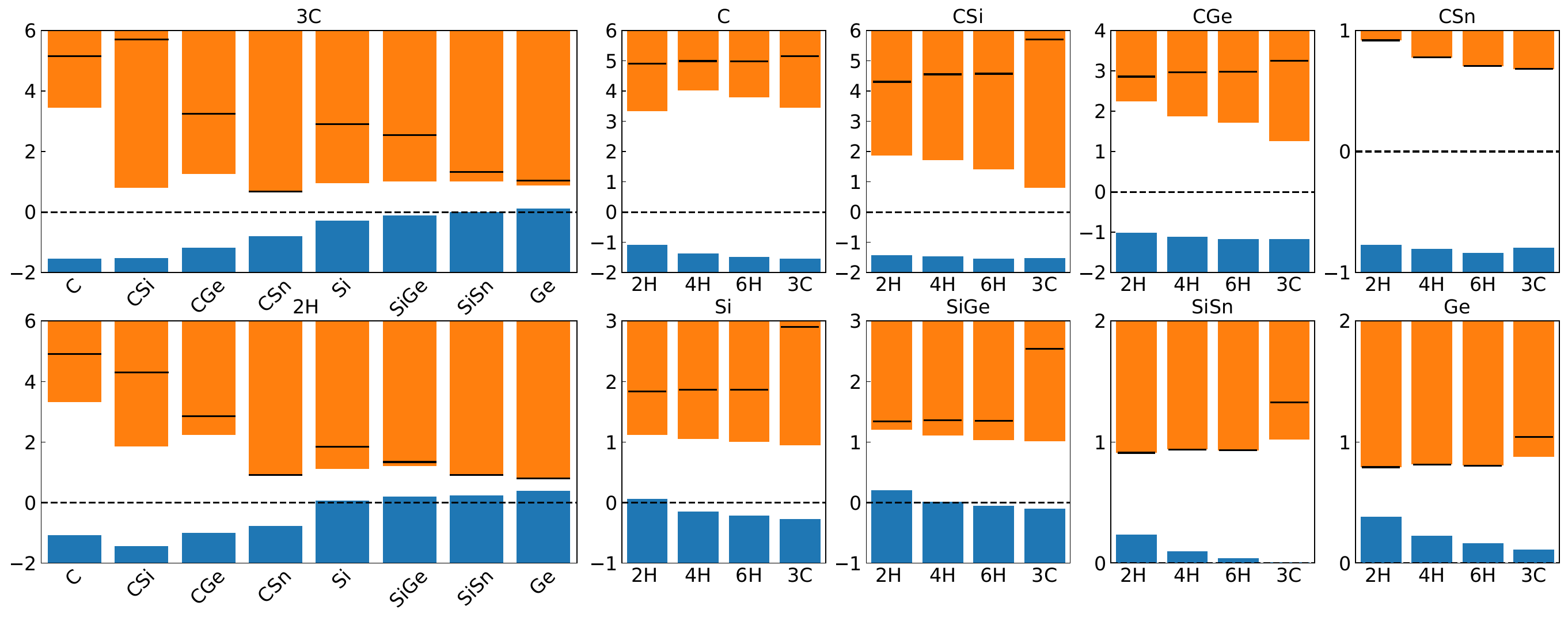}
    \caption{Relative valence and conduction band edge positions. Chemical trends with increasing atomic mass of atoms is presented in the left panel. Remaining panels show changes in band edge positions for different polytypes. Direct gaps are marked with black solid lines and branch point energies (which is set to 0) with dashed lines.} 
    \label{fig:Offset}
\end{figure*}

In order to predict relative positions of valence and conduction bands in different materials, we used the branch point energy approach (for details about the method, see Sec. \ref{Ch:method}). The chemical trend for different chemical compositions is presented on Fig. \ref{fig:Offset}. It can be observed that generally valence band moves up and conduction band moves down in energy with increasing average atomic mass, with more pronounced changes observed in the conduction band. On the other hand, both bands move down in energy, with some exceptions, with decreasing hexagonality. In materials with carbon, the biggest changes can be observed in the position of the conduction band, whereas for rest materials changes are mostly in valence band. For most materials the band alignment between different polytypes of the same material is of type II. Exceptions are carbon with 2H conduction band lying below other hexagonal phases and CSn with 3C valence band above 4H and 6H. Other exceptions are SiSn and Ge for which conduction bands moves to higher energies with decreasing hexagonality. The predicted branch point energies with respect to the valence band maximum are given in Tab. \ref{Tab:EgDeltas} and resulting band edge positions are comparable to those derived in previous works. \cite{SiGe,SiGe_offset}. 

\subsection{Discussion}\label{Ch:discussion}

When considering potential applications of the studied IV-IV materials in optoelectronics, lattice constants and band gaps are key, hence Fig. \ref{fig:EgDeltas} requires further analysis. Starting from the lowest lattice constant, we have carbon with a very wide band gap of 4.4-5.4 eV. In all polytypes studied in this work it is an indirect gap material, which impairs its light emitting properties. Nonetheless it can be used in ultraviolet detectors \cite{UVdet}. More common applications are related to its extreme mechanical properties and transparency in a wide spectral range. 

Next, we have CSi and CGe in a spectrum range of 2.3-3.3 eV. Band gaps in the visible spectrum range and good mechanical properties (bulk modulus of around 230 GPa and 200 GPa respectively, see Tab. \ref{Tab:lattice}) make them interesting materials for applications. Similarly to carbon, they have indirect band gaps, which limits their light emitting efficiency. Both materials seem to have comparable properties (lattice constant, bulk modulus, band gap, strong polytypism), which can be explained by similar radii of Si and Ge atoms. On the other hand, technological realisation of CGe is considerably more challenging than CSi.

CSn is an interesting case, as it is the only carbon containing  material with a direct band gap in all four polytypes. Its value of 1.5-1.7 eV lies on the border between visible and infrared light. Moreover, we calculated interband momentum matrix element of the fundamental transition and found it to be comparable to transitions in group III nitrides (see Fig. 2 in supplementary data). Direct band gap with big oscillator strength is a unique property in this material group, as the other direct gap materials (Ge and SiSn), have small momentum matrix elements in comparison to common direct gap semiconductors like GaAs. It is therefore very interesting for applications, although it is also predicted to be the most unstable from all studied materials, with formation energy per atom of 0.784 eV, which may present additional challenges. Nonetheless further advances in growth techniques of nanostructures may overcome such limitations.

Considering the great industrial importance of silicon it is important to study properties of its polytypes. In optoelectronics it is widely used as an active material in solar cells despite its indirect energy gap. Compared to the cubic phase, the 2H phase has a smaller energy gap of 1.058 eV. Folding of L point causes that also direct gap is much lower in $p$H than in 3C. The fact that the minimum in the cubic phase is next to the X point implies that this material also has an indirect band gap in the hexagonal phases. 

Among the studied materials, the most promising in terms of applications are Ge and SiGe with the band gaps in infrared spectrum. Because of the direct gap in Ge, the material is promising for light emitters. However, it is required to incorporate Si into the structure since pure 2H Ge, despite its direct gap, has a very weak photoluminescence. The reason for that lies in selection rules, which has been already discussed in Sec. \ref{Ch:results}. It was shown that incorporating even small amounts of Si relax the selection rules and enhance its light emitting properties. In fact the photoluminestence of nanowires made from the SiGe alloy have been shown to be comparable in strength to III-V direct semiconductors \cite{Ge_PL}. As Ge and Si have similar lattice constants, it is an attractive material system, as a wide range of alloy mixing can be achieved, which enhances tunability of the band gap. Another advantage is that it is lattice matched to silicon substrates.

SiSn system has somewhat similar properties to Ge. Just like Ge in hexagonal phases, SiSn has a direct gap with lowest transition forbidden by selection rules and an indirect gap in the cubic phase. It can be, therefore, expected that this system in the form of an alloy can be optically active, similarly to 2H Ge. As in the case of GeSn, alloying Si with Sn in the cubic form can make it a direct gap semiconductor, which was shown in low dimensional structures \cite{SiSn_ND}. It theoretically has great range of the band gap tunability, if one can overcomes the tight miscibility limits.

GeSn in another promising material. In our calculations where we treat it as compound, it has a close to 0 direct band gap. It is, therefore, expected that in an alloy range between pure Ge and GeSn the compound has a direct gap for hexagonal phases in the infrared part of spectrum. In the case of its cubic phase, Ge is an indirect gap semiconductor, so in order to make this material optically active in photoluminescence, certain amount of Sn is required, over 10\% \cite{GeSn_directGap,GeSn_DirectGap2,GeSn_DirectGap3}. The solubility limit is normally around 1\% but in nanowires this limit can be exceeded \cite{GeSn,GeSn2}. GeSn alloy is therefore an excellent candidate for tunable light emitters in the infrared spectrum.

\section{Conclusions}\label{Ch:conclusion}

In this work, we presented a comprehensive study of group IV semiconductors including its binaries in four polytypes 3C, 6H, 4H and 2H. Starting from analysis of their geometry, the most pronounced structural differences i.e. stacking of subsequent bilayers, and similarities i.e. similar in plane lattice constants, bulk modulus and nearest neighbors atomic configurations have been pointed out. To get insight in correlation of geometry and stability, we used the ANNNI model which has been previously used to explain properties of hexagonal and cubic materials. Materials with the strongest polytypic behaviour have been identified. Exhaustive analysis of electronic properties has been performed as well. The dominant differences in band structures arise from band folding and reduction of symmetry. Since the structure of six top valence bands is similar for hexagonal polytypes, a common six \kp formalism has been proposed for its description. The behaviour of band gaps as a function of hexagonality has been studied and for some materials the change in the type of band gap (direct vs. indirect) has been observed. Relative band edge positions between different materials have been derived by means of branch point energy. Such comparative studies of polytypes properties are crucial for better understanding and modeling of nanostructures, where many crystallographic phases can coexist. Finally, on the basis of the obtained results, we discuss all of the materials in terms of their potential applications in optoelectronics. In particular we found CSn, Ge, SiSn and GeSn materials to be the most promising as light emitters due to their direct band gap.

\section{Methods}\label{Ch:method}

Simulations in this work were done by means of density functional theory in plane waves representation as implemented in VASP code \cite{VASP,VASP2,VASP3,VASP_PAW}. Geometry optimization was performed with the LDA functional \cite{LDA1,LDA2} as it was shown to give the smallest deviation from experiment in lattice constants for theirs material group in the 3C phase \cite{LatticeCalc}. Bulk modulus have been derived by fitting energy vs. volume curve to the Birch–Murnaghan equation of state. For band structure calculations we used the MBJ \cite{MBJ} functional which is known to improve the description of band gaps over LDA, with accuracy comparable to more advanced methods as hybrid functionals or GW at the computational effort of meta-GGA \cite{MBJ_III-V,MBJ_III-N}. All calculations were performed with energy cutoff of 600eV and a $\Gamma$ centered Brillouin zone sampling of $20\times20\times Y$, where Y is 20, 12, 9, 6, 4 for 3C, 2H, 3C in hexagonal supercell, 4H, 6H respectively. The electronic energy convergence threshold was set to $10^{-8}$ eV and forces in equilibrium state were below $10^{-4}$ eV/\AA. The procedure of fitting six \kp model to DFT band structures are described in \cite{30kp_new}. The branch point energy calculations were performed by averaging $p$ lowest conduction and $2p$ highest valence bands of the $p$H polytype over Brillouin zone \cite{BP_approx}. This methodology was proposed as an alternative to the original Green-function method \cite{BP} and has been successfully used in predicting semiconductor branch point energies \cite{BP_approx}.

\begin{acknowledgments}
This work has been funded within the grant of the National
Science Center Poland (OPUS11, UMO-2016/21/B/ST7/01267).
Calculations have been carried out using resources provided by Wroclaw Centre for Networking and Supercomputing and Interdisciplinary Centre for Mathematical and Computational Modeling (ICM) at the University of Warsaw.
\end{acknowledgments}

\section*{Supplementary data}
Additional data not presented in this manuscript are available on figshare: \url{https://doi.org/10.6084/m9.figshare.25527007.v1}.

\bibliographystyle{aipnum4-1}
\bibliography{Ref}

\begin{thebibliography}{56}%
\makeatletter
\providecommand \@ifxundefined [1]{%
 \@ifx{#1\undefined}
}%
\providecommand \@ifnum [1]{%
 \ifnum #1\expandafter \@firstoftwo
 \else \expandafter \@secondoftwo
 \fi
}%
\providecommand \@ifx [1]{%
 \ifx #1\expandafter \@firstoftwo
 \else \expandafter \@secondoftwo
 \fi
}%
\providecommand \natexlab [1]{#1}%
\providecommand \enquote  [1]{``#1''}%
\providecommand \bibnamefont  [1]{#1}%
\providecommand \bibfnamefont [1]{#1}%
\providecommand \citenamefont [1]{#1}%
\providecommand \href@noop [0]{\@secondoftwo}%
\providecommand \href [0]{\begingroup \@sanitize@url \@href}%
\providecommand \@href[1]{\@@startlink{#1}\@@href}%
\providecommand \@@href[1]{\endgroup#1\@@endlink}%
\providecommand \@sanitize@url [0]{\catcode `\\12\catcode `\$12\catcode
  `\&12\catcode `\#12\catcode `\^12\catcode `\_12\catcode `\%12\relax}%
\providecommand \@@startlink[1]{}%
\providecommand \@@endlink[0]{}%
\providecommand \url  [0]{\begingroup\@sanitize@url \@url }%
\providecommand \@url [1]{\endgroup\@href {#1}{\urlprefix }}%
\providecommand \urlprefix  [0]{URL }%
\providecommand \Eprint [0]{\href }%
\providecommand \doibase [0]{http://dx.doi.org/}%
\providecommand \selectlanguage [0]{\@gobble}%
\providecommand \bibinfo  [0]{\@secondoftwo}%
\providecommand \bibfield  [0]{\@secondoftwo}%
\providecommand \translation [1]{[#1]}%
\providecommand \BibitemOpen [0]{}%
\providecommand \bibitemStop [0]{}%
\providecommand \bibitemNoStop [0]{.\EOS\space}%
\providecommand \EOS [0]{\spacefactor3000\relax}%
\providecommand \BibitemShut  [1]{\csname bibitem#1\endcsname}%
\let\auto@bib@innerbib\@empty
\bibitem [{\citenamefont {Wong‐Leung}\ \emph {et~al.}(2019)\citenamefont
  {Wong‐Leung}, \citenamefont {Yang}, \citenamefont {Li}, \citenamefont
  {Karuturi}, \citenamefont {Fu}, \citenamefont {Tan},\ and\ \citenamefont
  {Jagadish}}]{III-V_NW_review}%
  \BibitemOpen
  \bibfield  {author} {\bibinfo {author} {\bibfnamefont {J.}~\bibnamefont
  {Wong‐Leung}}, \bibinfo {author} {\bibfnamefont {I.}~\bibnamefont {Yang}},
  \bibinfo {author} {\bibfnamefont {Z.}~\bibnamefont {Li}}, \bibinfo {author}
  {\bibfnamefont {S.~K.}\ \bibnamefont {Karuturi}}, \bibinfo {author}
  {\bibfnamefont {L.}~\bibnamefont {Fu}}, \bibinfo {author} {\bibfnamefont
  {H.~H.}\ \bibnamefont {Tan}}, \ and\ \bibinfo {author} {\bibfnamefont
  {C.}~\bibnamefont {Jagadish}},\ }\href {\doibase 10.1002/adma.201904359}
  {\bibfield  {journal} {\bibinfo  {journal} {Advanced Materials}\ }\textbf
  {\bibinfo {volume} {32}} (\bibinfo {year} {2019}),\
  10.1002/adma.201904359}\BibitemShut {NoStop}%
\bibitem [{\citenamefont {Dasgupta}\ \emph {et~al.}(2014)\citenamefont
  {Dasgupta}, \citenamefont {Sun}, \citenamefont {Liu}, \citenamefont
  {Brittman}, \citenamefont {Andrews}, \citenamefont {Lim}, \citenamefont
  {Gao}, \citenamefont {Yan},\ and\ \citenamefont {Yang}}]{NW_review}%
  \BibitemOpen
  \bibfield  {author} {\bibinfo {author} {\bibfnamefont {N.~P.}\ \bibnamefont
  {Dasgupta}}, \bibinfo {author} {\bibfnamefont {J.}~\bibnamefont {Sun}},
  \bibinfo {author} {\bibfnamefont {C.}~\bibnamefont {Liu}}, \bibinfo {author}
  {\bibfnamefont {S.}~\bibnamefont {Brittman}}, \bibinfo {author}
  {\bibfnamefont {S.~C.}\ \bibnamefont {Andrews}}, \bibinfo {author}
  {\bibfnamefont {J.}~\bibnamefont {Lim}}, \bibinfo {author} {\bibfnamefont
  {H.}~\bibnamefont {Gao}}, \bibinfo {author} {\bibfnamefont {R.}~\bibnamefont
  {Yan}}, \ and\ \bibinfo {author} {\bibfnamefont {P.}~\bibnamefont {Yang}},\
  }\href {\doibase 10.1002/adma.201305929} {\bibfield  {journal} {\bibinfo
  {journal} {Advanced Materials}\ }\textbf {\bibinfo {volume} {26}},\ \bibinfo
  {pages} {2137} (\bibinfo {year} {2014})}\BibitemShut {NoStop}%
\bibitem [{\citenamefont {Nehra}\ \emph {et~al.}(2020)\citenamefont {Nehra},
  \citenamefont {Dilbaghi}, \citenamefont {Marrazza}, \citenamefont {Kaushik},
  \citenamefont {Abolhassani}, \citenamefont {Mishra}, \citenamefont {Kim},\
  and\ \citenamefont {Kumar}}]{NW_review2}%
  \BibitemOpen
  \bibfield  {author} {\bibinfo {author} {\bibfnamefont {M.}~\bibnamefont
  {Nehra}}, \bibinfo {author} {\bibfnamefont {N.}~\bibnamefont {Dilbaghi}},
  \bibinfo {author} {\bibfnamefont {G.}~\bibnamefont {Marrazza}}, \bibinfo
  {author} {\bibfnamefont {A.}~\bibnamefont {Kaushik}}, \bibinfo {author}
  {\bibfnamefont {R.}~\bibnamefont {Abolhassani}}, \bibinfo {author}
  {\bibfnamefont {Y.~K.}\ \bibnamefont {Mishra}}, \bibinfo {author}
  {\bibfnamefont {K.~H.}\ \bibnamefont {Kim}}, \ and\ \bibinfo {author}
  {\bibfnamefont {S.}~\bibnamefont {Kumar}},\ }\href {\doibase
  10.1016/j.nanoen.2020.104991} {\bibfield  {journal} {\bibinfo  {journal}
  {Nano Energy}\ }\textbf {\bibinfo {volume} {76}},\ \bibinfo {pages} {104991}
  (\bibinfo {year} {2020})}\BibitemShut {NoStop}%
\bibitem [{\citenamefont {Sun}\ \emph {et~al.}(2019)\citenamefont {Sun},
  \citenamefont {Dong}, \citenamefont {Yu}, \citenamefont {Xu},\ and\
  \citenamefont {Chen}}]{NW_review3}%
  \BibitemOpen
  \bibfield  {author} {\bibinfo {author} {\bibfnamefont {Y.}~\bibnamefont
  {Sun}}, \bibinfo {author} {\bibfnamefont {T.}~\bibnamefont {Dong}}, \bibinfo
  {author} {\bibfnamefont {L.}~\bibnamefont {Yu}}, \bibinfo {author}
  {\bibfnamefont {J.}~\bibnamefont {Xu}}, \ and\ \bibinfo {author}
  {\bibfnamefont {K.}~\bibnamefont {Chen}},\ }\href {\doibase
  10.1002/adma.201903945} {\bibfield  {journal} {\bibinfo  {journal} {Advanced
  Materials}\ }\textbf {\bibinfo {volume} {32}} (\bibinfo {year} {2019}),\
  10.1002/adma.201903945}\BibitemShut {NoStop}%
\bibitem [{\citenamefont {Amato}\ \emph {et~al.}(2013)\citenamefont {Amato},
  \citenamefont {Palummo}, \citenamefont {Rurali},\ and\ \citenamefont
  {Ossicini}}]{SiGe_NW}%
  \BibitemOpen
  \bibfield  {author} {\bibinfo {author} {\bibfnamefont {M.}~\bibnamefont
  {Amato}}, \bibinfo {author} {\bibfnamefont {M.}~\bibnamefont {Palummo}},
  \bibinfo {author} {\bibfnamefont {R.}~\bibnamefont {Rurali}}, \ and\ \bibinfo
  {author} {\bibfnamefont {S.}~\bibnamefont {Ossicini}},\ }\href {\doibase
  10.1021/cr400261y} {\bibfield  {journal} {\bibinfo  {journal} {Chemical
  Reviews}\ }\textbf {\bibinfo {volume} {114}},\ \bibinfo {pages} {1371–1412}
  (\bibinfo {year} {2013})}\BibitemShut {NoStop}%
\bibitem [{\citenamefont {McIntyre}\ and\ \citenamefont {Fontcuberta~i
  Morral}(2020)}]{NanoGrowth}%
  \BibitemOpen
  \bibfield  {author} {\bibinfo {author} {\bibfnamefont {P.}~\bibnamefont
  {McIntyre}}\ and\ \bibinfo {author} {\bibfnamefont {A.}~\bibnamefont
  {Fontcuberta~i Morral}},\ }\href {\doibase 10.1016/j.mtnano.2019.100058}
  {\bibfield  {journal} {\bibinfo  {journal} {Materials Today Nano}\ }\textbf
  {\bibinfo {volume} {9}},\ \bibinfo {pages} {100058} (\bibinfo {year}
  {2020})}\BibitemShut {NoStop}%
\bibitem [{\citenamefont {Fukata}\ and\ \citenamefont
  {Jevasuwan}(2024)}]{NanoGrowth2}%
  \BibitemOpen
  \bibfield  {author} {\bibinfo {author} {\bibfnamefont {N.}~\bibnamefont
  {Fukata}}\ and\ \bibinfo {author} {\bibfnamefont {W.}~\bibnamefont
  {Jevasuwan}},\ }\href {\doibase 10.1088/1361-6528/ad15b8} {\bibfield
  {journal} {\bibinfo  {journal} {Nanotechnology}\ }\textbf {\bibinfo {volume}
  {35}},\ \bibinfo {pages} {122001} (\bibinfo {year} {2024})}\BibitemShut
  {NoStop}%
\bibitem [{\citenamefont {Glas}(2006)}]{Crit_dimension}%
  \BibitemOpen
  \bibfield  {author} {\bibinfo {author} {\bibfnamefont {F.}~\bibnamefont
  {Glas}},\ }\href {\doibase 10.1103/physrevb.74.121302} {\bibfield  {journal}
  {\bibinfo  {journal} {Physical Review B}\ }\textbf {\bibinfo {volume} {74}}
  (\bibinfo {year} {2006}),\ 10.1103/physrevb.74.121302}\BibitemShut {NoStop}%
\bibitem [{\citenamefont {Amato}, \citenamefont {Palummo},\ and\ \citenamefont
  {Ossicini}(2009{\natexlab{a}})}]{SiGe_NW2}%
  \BibitemOpen
  \bibfield  {author} {\bibinfo {author} {\bibfnamefont {M.}~\bibnamefont
  {Amato}}, \bibinfo {author} {\bibfnamefont {M.}~\bibnamefont {Palummo}}, \
  and\ \bibinfo {author} {\bibfnamefont {S.}~\bibnamefont {Ossicini}},\ }\href
  {\doibase 10.1103/physrevb.79.201302} {\bibfield  {journal} {\bibinfo
  {journal} {Physical Review B}\ }\textbf {\bibinfo {volume} {79}} (\bibinfo
  {year} {2009}{\natexlab{a}}),\ 10.1103/physrevb.79.201302}\BibitemShut
  {NoStop}%
\bibitem [{\citenamefont {Amato}, \citenamefont {Palummo},\ and\ \citenamefont
  {Ossicini}(2009{\natexlab{b}})}]{SiGe_NW3}%
  \BibitemOpen
  \bibfield  {author} {\bibinfo {author} {\bibfnamefont {M.}~\bibnamefont
  {Amato}}, \bibinfo {author} {\bibfnamefont {M.}~\bibnamefont {Palummo}}, \
  and\ \bibinfo {author} {\bibfnamefont {S.}~\bibnamefont {Ossicini}},\ }\href
  {\doibase 10.1103/physrevb.80.235333} {\bibfield  {journal} {\bibinfo
  {journal} {Physical Review B}\ }\textbf {\bibinfo {volume} {80}} (\bibinfo
  {year} {2009}{\natexlab{b}}),\ 10.1103/physrevb.80.235333}\BibitemShut
  {NoStop}%
\bibitem [{\citenamefont {Barth}, \citenamefont {Seifner},\ and\ \citenamefont
  {Maldonado}(2020)}]{polytypes}%
  \BibitemOpen
  \bibfield  {author} {\bibinfo {author} {\bibfnamefont {S.}~\bibnamefont
  {Barth}}, \bibinfo {author} {\bibfnamefont {M.~S.}\ \bibnamefont {Seifner}},
  \ and\ \bibinfo {author} {\bibfnamefont {S.}~\bibnamefont {Maldonado}},\
  }\href {\doibase 10.1021/acs.chemmater.9b04471} {\bibfield  {journal}
  {\bibinfo  {journal} {Chemistry of Materials}\ }\textbf {\bibinfo {volume}
  {32}},\ \bibinfo {pages} {2703–2741} (\bibinfo {year} {2020})}\BibitemShut
  {NoStop}%
\bibitem [{\citenamefont {Galvão~Tizei}\ and\ \citenamefont
  {Amato}(2020)}]{polytypes2}%
  \BibitemOpen
  \bibfield  {author} {\bibinfo {author} {\bibfnamefont {L.~H.}\ \bibnamefont
  {Galvão~Tizei}}\ and\ \bibinfo {author} {\bibfnamefont {M.}~\bibnamefont
  {Amato}},\ }\href {\doibase 10.1140/epjb/e2019-100375-7} {\bibfield
  {journal} {\bibinfo  {journal} {The European Physical Journal B}\ }\textbf
  {\bibinfo {volume} {93}} (\bibinfo {year} {2020}),\
  10.1140/epjb/e2019-100375-7}\BibitemShut {NoStop}%
\bibitem [{\citenamefont {Cheung}(2006)}]{SiC_polytypes}%
  \BibitemOpen
  \bibfield  {author} {\bibinfo {author} {\bibfnamefont {R.}~\bibnamefont
  {Cheung}},\ }\href {https://books.google.pl/books?id=hJySnYNE3B0C} {\emph
  {\bibinfo {title} {Silicon Carbide Microelectromechanical Systems for Harsh
  Environments}}}\ (\bibinfo  {publisher} {Imperial College Press},\ \bibinfo
  {year} {2006})\BibitemShut {NoStop}%
\bibitem [{\citenamefont {Kimoto}\ and\ \citenamefont
  {Cooper}(2014)}]{SiC_book}%
  \BibitemOpen
  \bibfield  {author} {\bibinfo {author} {\bibfnamefont {T.}~\bibnamefont
  {Kimoto}}\ and\ \bibinfo {author} {\bibfnamefont {J.~A.}\ \bibnamefont
  {Cooper}},\ }\href@noop {} {\emph {\bibinfo {title} {Fundamentals of Silicon
  Carbide Technology: Growth, characterization, devices and applications}}}\
  (\bibinfo  {publisher} {Wiley-IEEE Press},\ \bibinfo {year}
  {2014})\BibitemShut {NoStop}%
\bibitem [{\citenamefont {Feng}(2023)}]{SiC_book2}%
  \BibitemOpen
  \bibfield  {author} {\bibinfo {author} {\bibfnamefont {Z.~C.}\ \bibnamefont
  {Feng}},\ }\href@noop {} {\emph {\bibinfo {title} {Handbook of Silicon
  Carbide Materials and devices}}}\ (\bibinfo  {publisher} {CRC Press},\
  \bibinfo {year} {2023})\BibitemShut {NoStop}%
\bibitem [{\citenamefont {Wu}\ \emph {et~al.}(2015)\citenamefont {Wu},
  \citenamefont {Zhou}, \citenamefont {Yue}, \citenamefont {Wei},\ and\
  \citenamefont {Pan}}]{SiC_nanostructures}%
  \BibitemOpen
  \bibfield  {author} {\bibinfo {author} {\bibfnamefont {R.}~\bibnamefont
  {Wu}}, \bibinfo {author} {\bibfnamefont {K.}~\bibnamefont {Zhou}}, \bibinfo
  {author} {\bibfnamefont {C.~Y.}\ \bibnamefont {Yue}}, \bibinfo {author}
  {\bibfnamefont {J.}~\bibnamefont {Wei}}, \ and\ \bibinfo {author}
  {\bibfnamefont {Y.}~\bibnamefont {Pan}},\ }\href {\doibase
  10.1016/j.pmatsci.2015.01.003} {\bibfield  {journal} {\bibinfo  {journal}
  {Progress in Materials Science}\ }\textbf {\bibinfo {volume} {72}},\ \bibinfo
  {pages} {1–60} (\bibinfo {year} {2015})}\BibitemShut {NoStop}%
\bibitem [{\citenamefont {Fadaly}\ \emph {et~al.}(2020)\citenamefont {Fadaly},
  \citenamefont {Dijkstra}, \citenamefont {Suckert}, \citenamefont {Ziss},
  \citenamefont {van Tilburg}, \citenamefont {Mao}, \citenamefont {Ren},
  \citenamefont {van Lange}, \citenamefont {Korzun}, \citenamefont {Kölling},\
  and\ \citenamefont {et~al.}}]{Ge_PL}%
  \BibitemOpen
  \bibfield  {author} {\bibinfo {author} {\bibfnamefont {E.~M.}\ \bibnamefont
  {Fadaly}}, \bibinfo {author} {\bibfnamefont {A.}~\bibnamefont {Dijkstra}},
  \bibinfo {author} {\bibfnamefont {J.~R.}\ \bibnamefont {Suckert}}, \bibinfo
  {author} {\bibfnamefont {D.}~\bibnamefont {Ziss}}, \bibinfo {author}
  {\bibfnamefont {M.~A.}\ \bibnamefont {van Tilburg}}, \bibinfo {author}
  {\bibfnamefont {C.}~\bibnamefont {Mao}}, \bibinfo {author} {\bibfnamefont
  {Y.}~\bibnamefont {Ren}}, \bibinfo {author} {\bibfnamefont {V.~T.}\
  \bibnamefont {van Lange}}, \bibinfo {author} {\bibfnamefont {K.}~\bibnamefont
  {Korzun}}, \bibinfo {author} {\bibfnamefont {S.}~\bibnamefont {Kölling}}, \
  and\ \bibinfo {author} {\bibnamefont {et~al.}},\ }\href {\doibase
  10.1038/s41586-020-2150-y} {\bibfield  {journal} {\bibinfo  {journal}
  {Nature}\ }\textbf {\bibinfo {volume} {580}},\ \bibinfo {pages} {205–209}
  (\bibinfo {year} {2020})}\BibitemShut {NoStop}%
\bibitem [{\citenamefont {Rödl}\ \emph {et~al.}(2019)\citenamefont {Rödl},
  \citenamefont {Furthmüller}, \citenamefont {Suckert}, \citenamefont
  {Armuzza}, \citenamefont {Bechstedt},\ and\ \citenamefont
  {Botti}}]{Ge_OscStr}%
  \BibitemOpen
  \bibfield  {author} {\bibinfo {author} {\bibfnamefont {C.}~\bibnamefont
  {Rödl}}, \bibinfo {author} {\bibfnamefont {J.}~\bibnamefont {Furthmüller}},
  \bibinfo {author} {\bibfnamefont {J.~R.}\ \bibnamefont {Suckert}}, \bibinfo
  {author} {\bibfnamefont {V.}~\bibnamefont {Armuzza}}, \bibinfo {author}
  {\bibfnamefont {F.}~\bibnamefont {Bechstedt}}, \ and\ \bibinfo {author}
  {\bibfnamefont {S.}~\bibnamefont {Botti}},\ }\href {\doibase
  10.1103/physrevmaterials.3.034602} {\bibfield  {journal} {\bibinfo  {journal}
  {Physical Review Materials}\ }\textbf {\bibinfo {volume} {3}} (\bibinfo
  {year} {2019}),\ 10.1103/physrevmaterials.3.034602}\BibitemShut {NoStop}%
\bibitem [{\citenamefont {Suckert}\ \emph {et~al.}(2021)\citenamefont
  {Suckert}, \citenamefont {Rödl}, \citenamefont {Furthmüller}, \citenamefont
  {Bechstedt},\ and\ \citenamefont {Botti}}]{Ge_strain}%
  \BibitemOpen
  \bibfield  {author} {\bibinfo {author} {\bibfnamefont {J.~R.}\ \bibnamefont
  {Suckert}}, \bibinfo {author} {\bibfnamefont {C.}~\bibnamefont {Rödl}},
  \bibinfo {author} {\bibfnamefont {J.}~\bibnamefont {Furthmüller}}, \bibinfo
  {author} {\bibfnamefont {F.}~\bibnamefont {Bechstedt}}, \ and\ \bibinfo
  {author} {\bibfnamefont {S.}~\bibnamefont {Botti}},\ }\href {\doibase
  10.1103/physrevmaterials.5.024602} {\bibfield  {journal} {\bibinfo  {journal}
  {Physical Review Materials}\ }\textbf {\bibinfo {volume} {5}} (\bibinfo
  {year} {2021}),\ 10.1103/physrevmaterials.5.024602}\BibitemShut {NoStop}%
\bibitem [{\citenamefont {Belabbes}, \citenamefont {Bechstedt},\ and\
  \citenamefont {Botti}(2022)}]{Ge_perturbations}%
  \BibitemOpen
  \bibfield  {author} {\bibinfo {author} {\bibfnamefont {A.}~\bibnamefont
  {Belabbes}}, \bibinfo {author} {\bibfnamefont {F.}~\bibnamefont {Bechstedt}},
  \ and\ \bibinfo {author} {\bibfnamefont {S.}~\bibnamefont {Botti}},\ }\href
  {\doibase 10.1002/pssr.202100555} {\bibfield  {journal} {\bibinfo  {journal}
  {physica status solidi (RRL) – Rapid Research Letters}\ }\textbf {\bibinfo
  {volume} {16}} (\bibinfo {year} {2022}),\ 10.1002/pssr.202100555}\BibitemShut
  {NoStop}%
\bibitem [{\citenamefont {Borlido}\ \emph {et~al.}(2021)\citenamefont
  {Borlido}, \citenamefont {Suckert}, \citenamefont {Furthmüller},
  \citenamefont {Bechstedt}, \citenamefont {Botti},\ and\ \citenamefont
  {Rödl}}]{Ge_alloy}%
  \BibitemOpen
  \bibfield  {author} {\bibinfo {author} {\bibfnamefont {P.}~\bibnamefont
  {Borlido}}, \bibinfo {author} {\bibfnamefont {J.~R.}\ \bibnamefont
  {Suckert}}, \bibinfo {author} {\bibfnamefont {J.}~\bibnamefont
  {Furthmüller}}, \bibinfo {author} {\bibfnamefont {F.}~\bibnamefont
  {Bechstedt}}, \bibinfo {author} {\bibfnamefont {S.}~\bibnamefont {Botti}}, \
  and\ \bibinfo {author} {\bibfnamefont {C.}~\bibnamefont {Rödl}},\ }\href
  {\doibase 10.1103/physrevmaterials.5.114604} {\bibfield  {journal} {\bibinfo
  {journal} {Physical Review Materials}\ }\textbf {\bibinfo {volume} {5}}
  (\bibinfo {year} {2021}),\ 10.1103/physrevmaterials.5.114604}\BibitemShut
  {NoStop}%
\bibitem [{\citenamefont {Polak}, \citenamefont {Scharoch},\ and\ \citenamefont
  {Kudrawiec}(2017)}]{GeSn_directGap}%
  \BibitemOpen
  \bibfield  {author} {\bibinfo {author} {\bibfnamefont {M.~P.}\ \bibnamefont
  {Polak}}, \bibinfo {author} {\bibfnamefont {P.}~\bibnamefont {Scharoch}}, \
  and\ \bibinfo {author} {\bibfnamefont {R.}~\bibnamefont {Kudrawiec}},\ }\href
  {\doibase 10.1088/1361-6463/aa67bf} {\bibfield  {journal} {\bibinfo
  {journal} {Journal of Physics D: Applied Physics}\ }\textbf {\bibinfo
  {volume} {50}},\ \bibinfo {pages} {195103} (\bibinfo {year}
  {2017})}\BibitemShut {NoStop}%
\bibitem [{\citenamefont {Freitas}\ \emph {et~al.}(2016)\citenamefont
  {Freitas}, \citenamefont {Furthmüller}, \citenamefont {Bechstedt},
  \citenamefont {Marques},\ and\ \citenamefont {Teles}}]{GeSn_DirectGap2}%
  \BibitemOpen
  \bibfield  {author} {\bibinfo {author} {\bibfnamefont {F.~L.}\ \bibnamefont
  {Freitas}}, \bibinfo {author} {\bibfnamefont {J.}~\bibnamefont
  {Furthmüller}}, \bibinfo {author} {\bibfnamefont {F.}~\bibnamefont
  {Bechstedt}}, \bibinfo {author} {\bibfnamefont {M.}~\bibnamefont {Marques}},
  \ and\ \bibinfo {author} {\bibfnamefont {L.~K.}\ \bibnamefont {Teles}},\
  }\href {\doibase 10.1063/1.4942971} {\bibfield  {journal} {\bibinfo
  {journal} {Applied Physics Letters}\ }\textbf {\bibinfo {volume} {108}}
  (\bibinfo {year} {2016}),\ 10.1063/1.4942971}\BibitemShut {NoStop}%
\bibitem [{\citenamefont {Zelazna}\ \emph {et~al.}(2015)\citenamefont
  {Zelazna}, \citenamefont {Polak}, \citenamefont {Scharoch}, \citenamefont
  {Serafinczuk}, \citenamefont {Gladysiewicz}, \citenamefont {Misiewicz},
  \citenamefont {Dekoster},\ and\ \citenamefont {Kudrawiec}}]{GeSn_DirectGap3}%
  \BibitemOpen
  \bibfield  {author} {\bibinfo {author} {\bibfnamefont {K.}~\bibnamefont
  {Zelazna}}, \bibinfo {author} {\bibfnamefont {M.~P.}\ \bibnamefont {Polak}},
  \bibinfo {author} {\bibfnamefont {P.}~\bibnamefont {Scharoch}}, \bibinfo
  {author} {\bibfnamefont {J.}~\bibnamefont {Serafinczuk}}, \bibinfo {author}
  {\bibfnamefont {M.}~\bibnamefont {Gladysiewicz}}, \bibinfo {author}
  {\bibfnamefont {J.}~\bibnamefont {Misiewicz}}, \bibinfo {author}
  {\bibfnamefont {J.}~\bibnamefont {Dekoster}}, \ and\ \bibinfo {author}
  {\bibfnamefont {R.}~\bibnamefont {Kudrawiec}},\ }\href {\doibase
  10.1063/1.4917236} {\bibfield  {journal} {\bibinfo  {journal} {Applied
  Physics Letters}\ }\textbf {\bibinfo {volume} {106}} (\bibinfo {year}
  {2015}),\ 10.1063/1.4917236}\BibitemShut {NoStop}%
\bibitem [{\citenamefont {Meng}\ \emph {et~al.}(2016)\citenamefont {Meng},
  \citenamefont {Fenrich}, \citenamefont {Braun}, \citenamefont {McVittie},
  \citenamefont {Marshall}, \citenamefont {Harris},\ and\ \citenamefont
  {McIntyre}}]{GeSn}%
  \BibitemOpen
  \bibfield  {author} {\bibinfo {author} {\bibfnamefont {A.~C.}\ \bibnamefont
  {Meng}}, \bibinfo {author} {\bibfnamefont {C.~S.}\ \bibnamefont {Fenrich}},
  \bibinfo {author} {\bibfnamefont {M.~R.}\ \bibnamefont {Braun}}, \bibinfo
  {author} {\bibfnamefont {J.~P.}\ \bibnamefont {McVittie}}, \bibinfo {author}
  {\bibfnamefont {A.~F.}\ \bibnamefont {Marshall}}, \bibinfo {author}
  {\bibfnamefont {J.~S.}\ \bibnamefont {Harris}}, \ and\ \bibinfo {author}
  {\bibfnamefont {P.~C.}\ \bibnamefont {McIntyre}},\ }\href {\doibase
  10.1021/acs.nanolett.6b03316} {\bibfield  {journal} {\bibinfo  {journal}
  {Nano Letters}\ }\textbf {\bibinfo {volume} {16}},\ \bibinfo {pages}
  {7521–7529} (\bibinfo {year} {2016})}\BibitemShut {NoStop}%
\bibitem [{\citenamefont {Assali}\ \emph {et~al.}(2017)\citenamefont {Assali},
  \citenamefont {Dijkstra}, \citenamefont {Li}, \citenamefont {Koelling},
  \citenamefont {Verheijen}, \citenamefont {Gagliano}, \citenamefont {von~den
  Driesch}, \citenamefont {Buca}, \citenamefont {Koenraad}, \citenamefont
  {Haverkort},\ and\ \citenamefont {et~al.}}]{GeSn2}%
  \BibitemOpen
  \bibfield  {author} {\bibinfo {author} {\bibfnamefont {S.}~\bibnamefont
  {Assali}}, \bibinfo {author} {\bibfnamefont {A.}~\bibnamefont {Dijkstra}},
  \bibinfo {author} {\bibfnamefont {A.}~\bibnamefont {Li}}, \bibinfo {author}
  {\bibfnamefont {S.}~\bibnamefont {Koelling}}, \bibinfo {author}
  {\bibfnamefont {M.~A.}\ \bibnamefont {Verheijen}}, \bibinfo {author}
  {\bibfnamefont {L.}~\bibnamefont {Gagliano}}, \bibinfo {author}
  {\bibfnamefont {N.}~\bibnamefont {von~den Driesch}}, \bibinfo {author}
  {\bibfnamefont {D.}~\bibnamefont {Buca}}, \bibinfo {author} {\bibfnamefont
  {P.~M.}\ \bibnamefont {Koenraad}}, \bibinfo {author} {\bibfnamefont {J.~E.}\
  \bibnamefont {Haverkort}}, \ and\ \bibinfo {author} {\bibnamefont {et~al.}},\
  }\href {\doibase 10.1021/acs.nanolett.6b04627} {\bibfield  {journal}
  {\bibinfo  {journal} {Nano Letters}\ }\textbf {\bibinfo {volume} {17}},\
  \bibinfo {pages} {1538–1544} (\bibinfo {year} {2017})}\BibitemShut
  {NoStop}%
\bibitem [{\citenamefont {Chuang}\ and\ \citenamefont {Chang}(1996)}]{kp1}%
  \BibitemOpen
  \bibfield  {author} {\bibinfo {author} {\bibfnamefont {S.~L.}\ \bibnamefont
  {Chuang}}\ and\ \bibinfo {author} {\bibfnamefont {C.~S.}\ \bibnamefont
  {Chang}},\ }\href {\doibase 10.1103/physrevb.54.2491} {\bibfield  {journal}
  {\bibinfo  {journal} {Physical Review B}\ }\textbf {\bibinfo {volume} {54}},\
  \bibinfo {pages} {2491–2504} (\bibinfo {year} {1996})}\BibitemShut
  {NoStop}%
\bibitem [{\citenamefont {Chuang}(1996)}]{kp2}%
  \BibitemOpen
  \bibfield  {author} {\bibinfo {author} {\bibfnamefont {S.}~\bibnamefont
  {Chuang}},\ }\href {\doibase 10.1109/3.538786} {\bibfield  {journal}
  {\bibinfo  {journal} {IEEE Journal of Quantum Electronics}\ }\textbf
  {\bibinfo {volume} {32}},\ \bibinfo {pages} {1791–1800} (\bibinfo {year}
  {1996})}\BibitemShut {NoStop}%
\bibitem [{\citenamefont {Birner}\ \emph {et~al.}(2007)\citenamefont {Birner},
  \citenamefont {Zibold}, \citenamefont {Andlauer}, \citenamefont {Kubis},
  \citenamefont {Sabathil}, \citenamefont {Trellakis},\ and\ \citenamefont
  {Vogl}}]{nextnano}%
  \BibitemOpen
  \bibfield  {author} {\bibinfo {author} {\bibfnamefont {S.}~\bibnamefont
  {Birner}}, \bibinfo {author} {\bibfnamefont {T.}~\bibnamefont {Zibold}},
  \bibinfo {author} {\bibfnamefont {T.}~\bibnamefont {Andlauer}}, \bibinfo
  {author} {\bibfnamefont {T.}~\bibnamefont {Kubis}}, \bibinfo {author}
  {\bibfnamefont {M.}~\bibnamefont {Sabathil}}, \bibinfo {author}
  {\bibfnamefont {A.}~\bibnamefont {Trellakis}}, \ and\ \bibinfo {author}
  {\bibfnamefont {P.}~\bibnamefont {Vogl}},\ }\href {\doibase
  10.1109/ted.2007.902871} {\bibfield  {journal} {\bibinfo  {journal} {IEEE
  Transactions on Electron Devices}\ }\textbf {\bibinfo {volume} {54}},\
  \bibinfo {pages} {2137–2142} (\bibinfo {year} {2007})}\BibitemShut
  {NoStop}%
\bibitem [{\citenamefont {Keller}\ \emph {et~al.}(2023)\citenamefont {Keller},
  \citenamefont {Belabbes}, \citenamefont {Furthmüller}, \citenamefont
  {Bechstedt},\ and\ \citenamefont {Botti}}]{SiGe}%
  \BibitemOpen
  \bibfield  {author} {\bibinfo {author} {\bibfnamefont {M.}~\bibnamefont
  {Keller}}, \bibinfo {author} {\bibfnamefont {A.}~\bibnamefont {Belabbes}},
  \bibinfo {author} {\bibfnamefont {J.}~\bibnamefont {Furthmüller}}, \bibinfo
  {author} {\bibfnamefont {F.}~\bibnamefont {Bechstedt}}, \ and\ \bibinfo
  {author} {\bibfnamefont {S.}~\bibnamefont {Botti}},\ }\href {\doibase
  10.1103/physrevmaterials.7.064601} {\bibfield  {journal} {\bibinfo  {journal}
  {Physical Review Materials}\ }\textbf {\bibinfo {volume} {7}} (\bibinfo
  {year} {2023}),\ 10.1103/physrevmaterials.7.064601}\BibitemShut {NoStop}%
\bibitem [{\citenamefont {Lawaetz}(1972)}]{stability}%
  \BibitemOpen
  \bibfield  {author} {\bibinfo {author} {\bibfnamefont {P.}~\bibnamefont
  {Lawaetz}},\ }\href {\doibase 10.1103/physrevb.5.4039} {\bibfield  {journal}
  {\bibinfo  {journal} {Physical Review B}\ }\textbf {\bibinfo {volume} {5}},\
  \bibinfo {pages} {4039–4045} (\bibinfo {year} {1972})}\BibitemShut
  {NoStop}%
\bibitem [{\citenamefont {Chelikowsky}\ and\ \citenamefont
  {Phillips}(1978)}]{stability2}%
  \BibitemOpen
  \bibfield  {author} {\bibinfo {author} {\bibfnamefont {J.~R.}\ \bibnamefont
  {Chelikowsky}}\ and\ \bibinfo {author} {\bibfnamefont {J.~C.}\ \bibnamefont
  {Phillips}},\ }\href {\doibase 10.1103/physrevb.17.2453} {\bibfield
  {journal} {\bibinfo  {journal} {Physical Review B}\ }\textbf {\bibinfo
  {volume} {17}},\ \bibinfo {pages} {2453–2477} (\bibinfo {year}
  {1978})}\BibitemShut {NoStop}%
\bibitem [{\citenamefont {Käckell}, \citenamefont {Wenzien},\ and\
  \citenamefont {Bechstedt}(1994)}]{SiC_energy}%
  \BibitemOpen
  \bibfield  {author} {\bibinfo {author} {\bibfnamefont {P.}~\bibnamefont
  {Käckell}}, \bibinfo {author} {\bibfnamefont {B.}~\bibnamefont {Wenzien}}, \
  and\ \bibinfo {author} {\bibfnamefont {F.}~\bibnamefont {Bechstedt}},\ }\href
  {\doibase 10.1103/physrevb.50.17037} {\bibfield  {journal} {\bibinfo
  {journal} {Physical Review B}\ }\textbf {\bibinfo {volume} {50}},\ \bibinfo
  {pages} {17037–17046} (\bibinfo {year} {1994})}\BibitemShut {NoStop}%
\bibitem [{\citenamefont {Bechstedt}\ \emph {et~al.}(1997)\citenamefont
  {Bechstedt}, \citenamefont {Käckell}, \citenamefont {Zywietz}, \citenamefont
  {Karch}, \citenamefont {Adolph}, \citenamefont {Tenelsen},\ and\
  \citenamefont {Furthmüller}}]{ANNNI_SiC1}%
  \BibitemOpen
  \bibfield  {author} {\bibinfo {author} {\bibfnamefont {F.}~\bibnamefont
  {Bechstedt}}, \bibinfo {author} {\bibfnamefont {P.}~\bibnamefont {Käckell}},
  \bibinfo {author} {\bibfnamefont {A.}~\bibnamefont {Zywietz}}, \bibinfo
  {author} {\bibfnamefont {K.}~\bibnamefont {Karch}}, \bibinfo {author}
  {\bibfnamefont {B.}~\bibnamefont {Adolph}}, \bibinfo {author} {\bibfnamefont
  {K.}~\bibnamefont {Tenelsen}}, \ and\ \bibinfo {author} {\bibfnamefont
  {J.}~\bibnamefont {Furthmüller}},\ }\href {\doibase
  10.1002/1521-3951(199707)202:1<35::aid-pssb35>3.0.co;2-8} {\bibfield
  {journal} {\bibinfo  {journal} {physica status solidi (b)}\ }\textbf
  {\bibinfo {volume} {202}},\ \bibinfo {pages} {35–62} (\bibinfo {year}
  {1997})}\BibitemShut {NoStop}%
\bibitem [{\citenamefont {Cheng}, \citenamefont {Needs},\ and\ \citenamefont
  {Heine}(1988)}]{ANNNI_SiC2}%
  \BibitemOpen
  \bibfield  {author} {\bibinfo {author} {\bibfnamefont {C.}~\bibnamefont
  {Cheng}}, \bibinfo {author} {\bibfnamefont {R.~J.}\ \bibnamefont {Needs}}, \
  and\ \bibinfo {author} {\bibfnamefont {V.}~\bibnamefont {Heine}},\ }\href
  {\doibase 10.1088/0022-3719/21/6/012} {\bibfield  {journal} {\bibinfo
  {journal} {Journal of Physics C: Solid State Physics}\ }\textbf {\bibinfo
  {volume} {21}},\ \bibinfo {pages} {1049–1063} (\bibinfo {year}
  {1988})}\BibitemShut {NoStop}%
\bibitem [{\citenamefont {Bechstedt}\ and\ \citenamefont
  {Belabbes}(2013)}]{III-VRev}%
  \BibitemOpen
  \bibfield  {author} {\bibinfo {author} {\bibfnamefont {F.}~\bibnamefont
  {Bechstedt}}\ and\ \bibinfo {author} {\bibfnamefont {A.}~\bibnamefont
  {Belabbes}},\ }\href {\doibase 10.1088/0953-8984/25/27/273201} {\bibfield
  {journal} {\bibinfo  {journal} {Journal of Physics: Condensed Matter}\
  }\textbf {\bibinfo {volume} {25}},\ \bibinfo {pages} {273201} (\bibinfo
  {year} {2013})}\BibitemShut {NoStop}%
\bibitem [{\citenamefont {Koster}\ \emph {et~al.}(1963)\citenamefont {Koster},
  \citenamefont {Dimmock}, \citenamefont {Wheeler},\ and\ \citenamefont
  {Statz}}]{Koster}%
  \BibitemOpen
  \bibfield  {author} {\bibinfo {author} {\bibfnamefont {G.~F.}\ \bibnamefont
  {Koster}}, \bibinfo {author} {\bibfnamefont {J.~O.}\ \bibnamefont {Dimmock}},
  \bibinfo {author} {\bibfnamefont {R.~G.}\ \bibnamefont {Wheeler}}, \ and\
  \bibinfo {author} {\bibfnamefont {H.}~\bibnamefont {Statz}},\ }\href@noop {}
  {\emph {\bibinfo {title} {Properties of the thirty two point groups}}}\
  (\bibinfo  {publisher} {MIT Press},\ \bibinfo {year} {1963})\BibitemShut
  {NoStop}%
\bibitem [{\citenamefont {De}\ and\ \citenamefont {Pryor}(2014)}]{C_Si_Ge}%
  \BibitemOpen
  \bibfield  {author} {\bibinfo {author} {\bibfnamefont {A.}~\bibnamefont
  {De}}\ and\ \bibinfo {author} {\bibfnamefont {C.~E.}\ \bibnamefont {Pryor}},\
  }\href {\doibase 10.1088/0953-8984/26/4/045801} {\bibfield  {journal}
  {\bibinfo  {journal} {Journal of Physics: Condensed Matter}\ }\textbf
  {\bibinfo {volume} {26}},\ \bibinfo {pages} {045801} (\bibinfo {year}
  {2014})}\BibitemShut {NoStop}%
\bibitem [{\citenamefont {Rödl}\ \emph {et~al.}(2015)\citenamefont {Rödl},
  \citenamefont {Sander}, \citenamefont {Bechstedt}, \citenamefont {Vidal},
  \citenamefont {Olsson}, \citenamefont {Laribi},\ and\ \citenamefont
  {Guillemoles}}]{Si}%
  \BibitemOpen
  \bibfield  {author} {\bibinfo {author} {\bibfnamefont {C.}~\bibnamefont
  {Rödl}}, \bibinfo {author} {\bibfnamefont {T.}~\bibnamefont {Sander}},
  \bibinfo {author} {\bibfnamefont {F.}~\bibnamefont {Bechstedt}}, \bibinfo
  {author} {\bibfnamefont {J.}~\bibnamefont {Vidal}}, \bibinfo {author}
  {\bibfnamefont {P.}~\bibnamefont {Olsson}}, \bibinfo {author} {\bibfnamefont
  {S.}~\bibnamefont {Laribi}}, \ and\ \bibinfo {author} {\bibfnamefont {J.-F.}\
  \bibnamefont {Guillemoles}},\ }\href {\doibase 10.1103/physrevb.92.045207}
  {\bibfield  {journal} {\bibinfo  {journal} {Physical Review B}\ }\textbf
  {\bibinfo {volume} {92}} (\bibinfo {year} {2015}),\
  10.1103/physrevb.92.045207}\BibitemShut {NoStop}%
\bibitem [{\citenamefont {Joannopoulos}\ and\ \citenamefont
  {Cohen}(1973)}]{Si_Ge}%
  \BibitemOpen
  \bibfield  {author} {\bibinfo {author} {\bibfnamefont {J.~D.}\ \bibnamefont
  {Joannopoulos}}\ and\ \bibinfo {author} {\bibfnamefont {M.~L.}\ \bibnamefont
  {Cohen}},\ }\href {\doibase 10.1103/physrevb.8.2733} {\bibfield  {journal}
  {\bibinfo  {journal} {Physical Review B}\ }\textbf {\bibinfo {volume} {8}},\
  \bibinfo {pages} {2733–2755} (\bibinfo {year} {1973})}\BibitemShut
  {NoStop}%
\bibitem [{\citenamefont {Kaewmaraya}, \citenamefont {Vincent},\ and\
  \citenamefont {Amato}(2017)}]{SiGe_offset}%
  \BibitemOpen
  \bibfield  {author} {\bibinfo {author} {\bibfnamefont {T.}~\bibnamefont
  {Kaewmaraya}}, \bibinfo {author} {\bibfnamefont {L.}~\bibnamefont {Vincent}},
  \ and\ \bibinfo {author} {\bibfnamefont {M.}~\bibnamefont {Amato}},\ }\href
  {\doibase 10.1021/acs.jpcc.6b12782} {\bibfield  {journal} {\bibinfo
  {journal} {The Journal of Physical Chemistry C}\ }\textbf {\bibinfo {volume}
  {121}},\ \bibinfo {pages} {5820–5828} (\bibinfo {year} {2017})}\BibitemShut
  {NoStop}%
\bibitem [{\citenamefont {Jia}, \citenamefont {Zheng},\ and\ \citenamefont
  {Huang}(2020)}]{UVdet}%
  \BibitemOpen
  \bibfield  {author} {\bibinfo {author} {\bibfnamefont {L.}~\bibnamefont
  {Jia}}, \bibinfo {author} {\bibfnamefont {W.}~\bibnamefont {Zheng}}, \ and\
  \bibinfo {author} {\bibfnamefont {F.}~\bibnamefont {Huang}},\ }\href
  {\doibase 10.1186/s43074-020-00022-w} {\bibfield  {journal} {\bibinfo
  {journal} {PhotoniX}\ }\textbf {\bibinfo {volume} {1}} (\bibinfo {year}
  {2020}),\ 10.1186/s43074-020-00022-w}\BibitemShut {NoStop}%
\bibitem [{\citenamefont {Lozac’h}\ \emph {et~al.}(2018)\citenamefont
  {Lozac’h}, \citenamefont {Švrček}, \citenamefont {Askari}, \citenamefont
  {Mariotti}, \citenamefont {Ohashi}, \citenamefont {Koganezawa}, \citenamefont
  {Miyadera},\ and\ \citenamefont {Matsubara}}]{SiSn_ND}%
  \BibitemOpen
  \bibfield  {author} {\bibinfo {author} {\bibfnamefont {M.}~\bibnamefont
  {Lozac’h}}, \bibinfo {author} {\bibfnamefont {V.}~\bibnamefont {Švrček}},
  \bibinfo {author} {\bibfnamefont {S.}~\bibnamefont {Askari}}, \bibinfo
  {author} {\bibfnamefont {D.}~\bibnamefont {Mariotti}}, \bibinfo {author}
  {\bibfnamefont {N.}~\bibnamefont {Ohashi}}, \bibinfo {author} {\bibfnamefont
  {T.}~\bibnamefont {Koganezawa}}, \bibinfo {author} {\bibfnamefont
  {T.}~\bibnamefont {Miyadera}}, \ and\ \bibinfo {author} {\bibfnamefont
  {K.}~\bibnamefont {Matsubara}},\ }\href {\doibase
  10.1016/j.mtener.2017.12.008} {\bibfield  {journal} {\bibinfo  {journal}
  {Materials Today Energy}\ }\textbf {\bibinfo {volume} {7}},\ \bibinfo {pages}
  {87–97} (\bibinfo {year} {2018})}\BibitemShut {NoStop}%
\bibitem [{\citenamefont {Kresse}\ and\ \citenamefont {Hafner}(1993)}]{VASP}%
  \BibitemOpen
  \bibfield  {author} {\bibinfo {author} {\bibfnamefont {G.}~\bibnamefont
  {Kresse}}\ and\ \bibinfo {author} {\bibfnamefont {J.}~\bibnamefont
  {Hafner}},\ }\href {\doibase 10.1103/PhysRevB.47.558} {\bibfield  {journal}
  {\bibinfo  {journal} {Phys. Rev. B}\ }\textbf {\bibinfo {volume} {47}},\
  \bibinfo {pages} {558} (\bibinfo {year} {1993})}\BibitemShut {NoStop}%
\bibitem [{\citenamefont {Kresse}\ and\ \citenamefont
  {Furthmüller}(1996{\natexlab{a}})}]{VASP2}%
  \BibitemOpen
  \bibfield  {author} {\bibinfo {author} {\bibfnamefont {G.}~\bibnamefont
  {Kresse}}\ and\ \bibinfo {author} {\bibfnamefont {J.}~\bibnamefont
  {Furthmüller}},\ }\href {\doibase 10.1103/physrevb.54.11169} {\bibfield
  {journal} {\bibinfo  {journal} {Phys. Rev. B}\ }\textbf {\bibinfo {volume}
  {54}},\ \bibinfo {pages} {11169–11186} (\bibinfo {year}
  {1996}{\natexlab{a}})}\BibitemShut {NoStop}%
\bibitem [{\citenamefont {Kresse}\ and\ \citenamefont
  {Furthmüller}(1996{\natexlab{b}})}]{VASP3}%
  \BibitemOpen
  \bibfield  {author} {\bibinfo {author} {\bibfnamefont {G.}~\bibnamefont
  {Kresse}}\ and\ \bibinfo {author} {\bibfnamefont {J.}~\bibnamefont
  {Furthmüller}},\ }\href {\doibase
  https://doi.org/10.1016/0927-0256(96)00008-0} {\bibfield  {journal} {\bibinfo
   {journal} {Computational Materials Science}\ }\textbf {\bibinfo {volume}
  {6}},\ \bibinfo {pages} {15} (\bibinfo {year}
  {1996}{\natexlab{b}})}\BibitemShut {NoStop}%
\bibitem [{\citenamefont {Kresse}\ and\ \citenamefont
  {Joubert}(1999)}]{VASP_PAW}%
  \BibitemOpen
  \bibfield  {author} {\bibinfo {author} {\bibfnamefont {G.}~\bibnamefont
  {Kresse}}\ and\ \bibinfo {author} {\bibfnamefont {D.}~\bibnamefont
  {Joubert}},\ }\href {\doibase 10.1103/physrevb.59.1758} {\bibfield  {journal}
  {\bibinfo  {journal} {Phys. Rev. B}\ }\textbf {\bibinfo {volume} {59}},\
  \bibinfo {pages} {1758–1775} (\bibinfo {year} {1999})}\BibitemShut
  {NoStop}%
\bibitem [{\citenamefont {Ceperley}\ and\ \citenamefont {Alder}(1980)}]{LDA1}%
  \BibitemOpen
  \bibfield  {author} {\bibinfo {author} {\bibfnamefont {D.~M.}\ \bibnamefont
  {Ceperley}}\ and\ \bibinfo {author} {\bibfnamefont {B.~J.}\ \bibnamefont
  {Alder}},\ }\href {\doibase 10.1103/physrevlett.45.566} {\bibfield  {journal}
  {\bibinfo  {journal} {Physical Review Letters}\ }\textbf {\bibinfo {volume}
  {45}},\ \bibinfo {pages} {566–569} (\bibinfo {year} {1980})}\BibitemShut
  {NoStop}%
\bibitem [{\citenamefont {Perdew}\ and\ \citenamefont {Zunger}(1981)}]{LDA2}%
  \BibitemOpen
  \bibfield  {author} {\bibinfo {author} {\bibfnamefont {J.~P.}\ \bibnamefont
  {Perdew}}\ and\ \bibinfo {author} {\bibfnamefont {A.}~\bibnamefont
  {Zunger}},\ }\href {\doibase 10.1103/physrevb.23.5048} {\bibfield  {journal}
  {\bibinfo  {journal} {Physical Review B}\ }\textbf {\bibinfo {volume} {23}},\
  \bibinfo {pages} {5048–5079} (\bibinfo {year} {1981})}\BibitemShut
  {NoStop}%
\bibitem [{\citenamefont {Haas}, \citenamefont {Tran},\ and\ \citenamefont
  {Blaha}(2009)}]{LatticeCalc}%
  \BibitemOpen
  \bibfield  {author} {\bibinfo {author} {\bibfnamefont {P.}~\bibnamefont
  {Haas}}, \bibinfo {author} {\bibfnamefont {F.}~\bibnamefont {Tran}}, \ and\
  \bibinfo {author} {\bibfnamefont {P.}~\bibnamefont {Blaha}},\ }\href
  {\doibase 10.1103/PhysRevB.79.085104} {\bibfield  {journal} {\bibinfo
  {journal} {Phys. Rev. B}\ }\textbf {\bibinfo {volume} {79}},\ \bibinfo
  {pages} {085104} (\bibinfo {year} {2009})}\BibitemShut {NoStop}%
\bibitem [{\citenamefont {Tran}\ and\ \citenamefont {Blaha}(2009)}]{MBJ}%
  \BibitemOpen
  \bibfield  {author} {\bibinfo {author} {\bibfnamefont {F.}~\bibnamefont
  {Tran}}\ and\ \bibinfo {author} {\bibfnamefont {P.}~\bibnamefont {Blaha}},\
  }\href {\doibase 10.1103/PhysRevLett.102.226401} {\bibfield  {journal}
  {\bibinfo  {journal} {Phys. Rev. Lett.}\ }\textbf {\bibinfo {volume} {102}},\
  \bibinfo {pages} {226401} (\bibinfo {year} {2009})}\BibitemShut {NoStop}%
\bibitem [{\citenamefont {Kim}\ \emph {et~al.}(2010)\citenamefont {Kim},
  \citenamefont {Marsman}, \citenamefont {Kresse}, \citenamefont {Tran},\ and\
  \citenamefont {Blaha}}]{MBJ_III-V}%
  \BibitemOpen
  \bibfield  {author} {\bibinfo {author} {\bibfnamefont {Y.-S.}\ \bibnamefont
  {Kim}}, \bibinfo {author} {\bibfnamefont {M.}~\bibnamefont {Marsman}},
  \bibinfo {author} {\bibfnamefont {G.}~\bibnamefont {Kresse}}, \bibinfo
  {author} {\bibfnamefont {F.}~\bibnamefont {Tran}}, \ and\ \bibinfo {author}
  {\bibfnamefont {P.}~\bibnamefont {Blaha}},\ }\href {\doibase
  10.1103/PhysRevB.82.205212} {\bibfield  {journal} {\bibinfo  {journal} {Phys.
  Rev. B}\ }\textbf {\bibinfo {volume} {82}},\ \bibinfo {pages} {205212}
  (\bibinfo {year} {2010})}\BibitemShut {NoStop}%
\bibitem [{\citenamefont {Araujo}, \citenamefont {de~Almeida},\ and\
  \citenamefont {Ferreira~da Silva}(2013)}]{MBJ_III-N}%
  \BibitemOpen
  \bibfield  {author} {\bibinfo {author} {\bibfnamefont {R.~B.}\ \bibnamefont
  {Araujo}}, \bibinfo {author} {\bibfnamefont {J.~S.}\ \bibnamefont
  {de~Almeida}}, \ and\ \bibinfo {author} {\bibfnamefont {A.}~\bibnamefont
  {Ferreira~da Silva}},\ }\href {\doibase 10.1063/1.4829674} {\bibfield
  {journal} {\bibinfo  {journal} {Journal of Applied Physics}\ }\textbf
  {\bibinfo {volume} {114}},\ \bibinfo {pages} {183702} (\bibinfo {year}
  {2013})},\ \Eprint {http://arxiv.org/abs/https://doi.org/10.1063/1.4829674}
  {https://doi.org/10.1063/1.4829674} \BibitemShut {NoStop}%
\bibitem [{\citenamefont {Gawarecki}\ \emph {et~al.}(2022)\citenamefont
  {Gawarecki}, \citenamefont {Scharoch}, \citenamefont
  {Wi\ifmmode~\acute{s}\else \'{s}\fi{}niewski}, \citenamefont {Ziembicki},
  \citenamefont {M\k{a}czko}, \citenamefont {G\l{}adysiewicz},\ and\
  \citenamefont {Kudrawiec}}]{30kp_new}%
  \BibitemOpen
  \bibfield  {author} {\bibinfo {author} {\bibfnamefont {K.}~\bibnamefont
  {Gawarecki}}, \bibinfo {author} {\bibfnamefont {P.}~\bibnamefont {Scharoch}},
  \bibinfo {author} {\bibfnamefont {M.}~\bibnamefont {Wi\ifmmode~\acute{s}\else
  \'{s}\fi{}niewski}}, \bibinfo {author} {\bibfnamefont {J.}~\bibnamefont
  {Ziembicki}}, \bibinfo {author} {\bibfnamefont {H.~S.}\ \bibnamefont
  {M\k{a}czko}}, \bibinfo {author} {\bibfnamefont {M.}~\bibnamefont
  {G\l{}adysiewicz}}, \ and\ \bibinfo {author} {\bibfnamefont {R.}~\bibnamefont
  {Kudrawiec}},\ }\href {\doibase 10.1103/PhysRevB.105.045202} {\bibfield
  {journal} {\bibinfo  {journal} {Phys. Rev. B}\ }\textbf {\bibinfo {volume}
  {105}},\ \bibinfo {pages} {045202} (\bibinfo {year} {2022})}\BibitemShut
  {NoStop}%
\bibitem [{\citenamefont {Schleife}\ \emph {et~al.}(2009)\citenamefont
  {Schleife}, \citenamefont {Fuchs}, \citenamefont {Rödl}, \citenamefont
  {Furthmüller},\ and\ \citenamefont {Bechstedt}}]{BP_approx}%
  \BibitemOpen
  \bibfield  {author} {\bibinfo {author} {\bibfnamefont {A.}~\bibnamefont
  {Schleife}}, \bibinfo {author} {\bibfnamefont {F.}~\bibnamefont {Fuchs}},
  \bibinfo {author} {\bibfnamefont {C.}~\bibnamefont {Rödl}}, \bibinfo
  {author} {\bibfnamefont {J.}~\bibnamefont {Furthmüller}}, \ and\ \bibinfo
  {author} {\bibfnamefont {F.}~\bibnamefont {Bechstedt}},\ }\href {\doibase
  10.1063/1.3059569} {\bibfield  {journal} {\bibinfo  {journal} {Applied
  Physics Letters}\ }\textbf {\bibinfo {volume} {94}} (\bibinfo {year}
  {2009}),\ 10.1063/1.3059569}\BibitemShut {NoStop}%
\bibitem [{\citenamefont {Tersoff}(1984)}]{BP}%
  \BibitemOpen
  \bibfield  {author} {\bibinfo {author} {\bibfnamefont {J.}~\bibnamefont
  {Tersoff}},\ }\href {\doibase 10.1103/physrevb.30.4874} {\bibfield  {journal}
  {\bibinfo  {journal} {Physical Review B}\ }\textbf {\bibinfo {volume} {30}},\
  \bibinfo {pages} {4874–4877} (\bibinfo {year} {1984})}\BibitemShut
  {NoStop}%
\end{thebibliography}%

\end{document}